\documentclass[11pt,twoside,letterpaper,onecolumn,draftcls]{IEEEtran}
\usepackage{times,graphics,amssymb,amsmath,bm,cite}
\usepackage[dvips]{graphicx}
\usepackage[active]{srcltx}  
\usepackage{epsfig,latexsym}

\usepackage{times,graphics,amssymb,array,amsmath,bm,cite}
\usepackage[dvips]{graphicx}
\usepackage{booktabs}
\usepackage[active]{srcltx}  
\usepackage{epsfig,latexsym}

\newcommand{\eeq}{\end{equation}}

\newcommand{\br}{\mbox{\boldmath $r$}}

\newcommand{\bs}{\mbox{\boldmath $s$}}

\newcommand{\bq}{\mbox{\boldmath $q$}}
\newcommand{\bp}{\mbox{\boldmath $p$}}

\newcommand{\bM}{\mbox{\boldmath $M$}}

\newcommand{\bDelta}{\mbox{\boldmath $\Delta$}}

\newcommand{\bA}{\mbox{\boldmath $A$}}
\newcommand{\bP}{\mbox{\boldmath $P$}}

\newcommand{\bX}{\mbox{\boldmath $X$}}

\newcommand{\bS}{\mbox{\boldmath $S$}}

\newcommand{\blambda}{\mbox{\boldmath $\lambda$}}
\newcommand{\bZ}{\mbox{\boldmath $Z$}}

\newcommand{\bV}{\mbox{\boldmath $V$}}

\newcommand{\bLambda}{{\bf \Lambda}}

\newcommand{\bx}{\mbox{\boldmath $x$}}

\newcommand{\bn}{\mbox{\boldmath $n$}}
\newcommand{\bv}{\mbox{\boldmath $v$}}
\newcommand{\bd}{\mbox{\boldmath $d$}}

\newcommand{\bI}{\mbox{\boldmath $I$}}

\newcommand{\bSigma}{{\bf \Sigma}}

\newcommand{\bD}{\mbox{\boldmath $D$}}

\newcommand{\ds}{\displaystyle}
\newcommand{\bw}{\mbox{\boldmath $w$}}

\newcommand{\G}{{\cal G}}
\newcommand{\N}{{\cal N}_0}
\newcommand{\beq}{\begin{equation}}

\newtheorem{theorem}{Theorem} \newtheorem{lemma}{Lemma}
\def\QED{\mbox{\rule[0pt]{1.5ex}{1.5ex}}}
\def\proof{\noindent {\it Proof: }}
\def\endproof{\hspace*{\fill}~\QED\par\endtrivlist\unskip}

\begin{document}


\title{Spreading Code and  Widely-Linear Receiver Design: Non-Cooperative Games for Wireless CDMA Networks}
\author{{Stefano Buzzi, {\em Senior Member, IEEE},  H. Vincent Poor, {\em Fellow, IEEE}, and\\Alessio Zappone, {\em Student Member, IEEE}}\\
\thanks{The research was supported in part by the U. S. National
Science Foundation under Grant CNS-06-25637.
This paper was partly presented at the 2009 IEEE Sarnoff Symposium, Princeton, NJ, March 2009, and at the Int. Conference on Game Theory for Networks, Istanbul (Turkey), May 2009.
Stefano Buzzi and Alessio Zappone are with DAEIMI, University of Cassino, Via G. Di Biasio, 43, I-03043 Cassino (FR), Italy (e-mail: buzzi@unicas.it, alessio.zappone@unicas.it);
H. Vincent Poor is with the School of Engineering and Applied Science, Princeton University, Princeton, NJ, 08544, USA (e-mail: poor@princeton.edu). }
} \maketitle



\begin{abstract}
The issue of non-cooperative transceiver optimization in the uplink of a multiuser wireless code division multiple access data network with widely-linear detection at the receiver is considered. While previous work in this area has focused on a simple real signal model, in this paper a baseband complex representation of the data is used, so as to properly take into account the I and Q components of the received signal. For the case in which the received signal is improper, a widely-linear reception structure, processing separately the data and their complex conjugates, is considered.
Several non-cooperative resource allocation games are considered for this new scenario, and the performance gains granted by the use of widely-linear detection are assessed through theoretical analysis.
Numerical results confirm the validity of the theoretical findings, and show that exploiting the improper nature of the data in non-cooperative resource allocation brings remarkable performance improvements in multiuser wireless systems.


\end{abstract}
%

\section{Introduction, work motivation and summary of contributions}
In recent years, considerable attention has been given to the issue of joint transmitter and receiver optimization in wireless communication systems. Indeed, many studies have shown that coupling an optimized receiver structure with the use of judiciously designed signals at the transmitter can result in substantial performance improvements with respect to the case in which transmit signal optimization is not taken into account.

With reference to direct-sequence/code-division-multiple-access (DS/CDMA) systems, the leading air-interface technology of current third-generation wireless data networks, transmitter optimization essentially amounts to the problem of spreading code allocation among users.
This has been for years and continues to be a very active research area. Among the pioneering contributions in this field we find \cite{rupf}, where the sum-capacity maximizing sequences for an equal power synchronous DS/CDMA system are derived. The results of this study are then reconsidered and generalized in several other papers. In reference \cite{optimal}, as an example, the equal power constraint of \cite{rupf} is removed, and
optimal sequences for sum-capacity maximization in a synchronous system with arbitrary power profile are derived. The paper \cite{optimal0}
considers instead the problem of sum-power minimization under pre-assigned quality of service (QoS) constraints, and
shows that, when the sum-power minimizing signatures are used, the multiuser linear minimum mean square error (MMSE) receivers reduce to conventional matched filters. In
\cite{ulukusyates,ulukusyener,ensuring,rose}, instead, iterative algorithms that converge to these optimal spreading sequences are given, and it is also shown that maximization of the sum capacity is equivalent to minimization of the total squared correlation and to the minimization of the global (sum) mean square error (MSE). Similar results are also reported in \cite{rose2}. The papers \cite{concha,popescurose,honig}, instead, generalize part of these results to the case in which the multiuser system is affected by multipath fading.
While the above cited papers consider the problem of spreading code design aimed at optimizing a global performance measure, such as the sum-capacity, the sum-power, and the global MSE, the focus of this paper is on  competitive (i.e., non-cooperative) joint spreading code and receiver optimization, wherein each user selfishly chooses its own spreading code aimed at optimization of an individual performance measure, such as, for instance, the MSE incurred by that user.
Game theory \cite{gtbook}, a mathematical theory suited to describe the interactions among entities with contrasting interests, has emerged in the last decade as a natural tool to study such distributed resource allocation algorithms \cite{gt,rodriguez}, since it provides a natural framework for the design and  analysis of non-cooperative behavior. Indeed, the selfish users' interaction in a wireless network can be modeled as a non-cooperative game
in which the users' terminals are the players of the game tuning parameters such as transmit power and adopted spreading code  in order to compete
for network resources and to maximize a given utility measure. Any action
taken by a user, i.e. the choice of the waveform to transmit and of the radiated power, modifies the multiuser interference scenario in the network, and thus affects the performance of other users in the network.
Game theory is the natural tool for studying this kind of interaction.

A game-theoretic framework for non-cooperative energy efficiency maximization has been widely applied in the recent past to design resource allocation policies for DS/CDMA systems \cite{nara2,meshkati,buzzijsac}, ultrawideband (UWB) systems \cite{bacci}, and multiuser
multiple-input multiple-output (MIMO) systems \cite{eusipco2008}. In particular, the paper \cite{nara2} focuses on the problem of power control for non-cooperative energy efficiency maximization in a CDMA system. It is shown that the considered problem admits a unique Nash equilibrium (NE), and, also, pricing is used to force the NE point to be Pareto-efficient. The results of \cite{nara2} are extended in \cite{bacci} to the case of a multiuser UWB system subject to a multipath fading channel, and in \cite{meshkati} to the case in which energy efficiency is maximized in a multiuser CDMA system with respect to both transmit power and choice of the uplink linear receiver. Putting together the results of \cite{meshkati} with those regarding the issue of spreading code optimization  \cite{ulukusyates,ulukusyener,rose,ensuring}, in \cite{buzzijsac} the problem of energy efficiency maximization with respect to the choice of the transmit power, uplink linear receiver and transmit spreading code is addressed; it is shown here that the considered problem admits a unique NE point, that, remarkably, is also Pareto-efficient under the assumption that the number of active users does not exceed the processing gain.
Similar results are also reported in \cite{eusipco2008} with reference to a multiuser MIMO system, while the problem of adaptive implementation of the games developed in \cite{buzzijsac} is considered in the study \cite{buzzieurasip2009}.

All of these works, however, consider, for the sake of simplicity, a real signal model. Otherwise stated, I and Q components are not taken into account; while at a first look it might seem that the extension of the reported results to the complex case is straightforward, this is not always the case, especially when considering issues regarding the problem of spreading code optimization, such as for instance those reported in \cite{ulukusyates,ensuring}.

On the other hand, it is known that if modulations such as
binary phase shift keying (BPSK),
multilevel
amplitude-shift-keying (ASK), Gaussian minimum-shift-keying
(GMSK), and offset quadrature-amplitude-modulation (QAM) are used, receiver performance can be improved through the use of widely-linear (WL) reception structures, i.e. receivers that separately process the data and their conjugates \cite{Buzzi,BuzziLops,Yoon,lampe1,lampe2}. Indeed, when the modulations cited above are employed, the baseband equivalent of the data is an improper complex random process\footnote{A complex
random process $z(t)$ is said to be proper if its
pseudoautocorrelation function $R_{z}(t,u)=E\left\{z(t)
z(u)\right\}$ ($E\{\cdot\}$ denotes the statistical expectation) is
zero $\forall t, u$ while is said to be improper when $R_{z}(t,u)$
is non-zero.}.

This paper is thus devoted to the problem of non-cooperative resource allocation in a multiuser DS/CDMA wireless data network coupled with WL filtering. In particular, we consider the non-cooperative maximum-SINR game for spreading code allocation and WL receiver choice. A distributed iterative algorithm is proposed for signature and receiver update. This algorithm is shown to be convergent and to admit one unique stable fixed point. Several properties of this algorithm are also shown, mainly generalizing to the case of WL filtering and complex signal model many results scattered in references \cite{optimal,ulukusyates,ulukusyener,rose,ensuring}.
A non-cooperative maximum energy-efficiency game is then considered, showing that also in this case the use of WL filtering brings a substantial performance improvement with respect to the case in which linear detection is used at the receiver side. The analysis of this game is also carried out through large system analysis (LSA) arguments.

The rest of this paper is organized as follows. The next section contains the system model and a brief review of linear and WL detection rules. Section III is devoted to the problem of optimal spreading code allocation for non-cooperative SINR maximization: an iterative algorithm is proposed which is shown to admit one unique stable fixed point, corresponding to the global minimizer of the total mean square error. Section IV deals with the problem of non-cooperative energy efficiency maximization, assuming that also the transmit power is among the parameters to be tuned. Section V contains the numerical results, while concluding remarks are given in Section VI.

\section{System model and multiuser detection structure}
Consider the uplink of a $K$-user synchronous, single-cell, direct-sequence (DS)/CDMA  network with processing gain $N$ and subject to flat fading. In each symbol interval, the received signal is chip-matched filtered and sampled at the chip-rate, yielding the $N$-dimensional received vector
\beq
\br=\sum_{k=1}^{K}\sqrt{p_k}h_k e^{j\varphi_k}b_k\bs_k+\bn \; ,
\label{eq:r}
\eeq
where $p_k$ is the transmit power of the $k$-th user, $b_k\in\{-1,1\}$ is the information symbol of the $k$-th user, $h_k e^{j\varphi_k}$ is the complex channel gain, and $\bs_k$ is the spreading code of the $k$-th user. In the sequel we assume that $\bs_k$ is constrained to have unit norm. Finally, $\bn$ is the thermal noise, assumed to be a zero-mean white Gaussian random process with covariance matrix $2\N\bI_N$, with $\bI_N$ the identity matrix of order $N$.

Now, a linear detection structure decides on the symbol $b_k$ according to the rule
\beq
\widehat{b}_k=\mbox{sign}\left\{\Re\left[\bd_k^H \br\right]\right\} \; ,
\label{eq:decrule}
\eeq
with $(\cdot)^H$ denoting conjugate-transpose, $\widehat{b}_k$ the estimate of $b_k$,  $\bd_k$ the $N$-dimensional vector representing the receive filter for the user $k$ and $\Re[\cdot]$ denoting real part. The signal-to-interference-plus-noise ratio (SINR) $\gamma_k$ corresponding to the decision rule (\ref{eq:decrule}) can be written as
\beq
\gamma_k=\ds \frac{p_k h_k^2 |\bd_k^H \bs_k|^2}{2\N\|\bd_k\|^2 + \ds \sum_{i \neq k} p_i h_i^2
|\bd_k^H \bs_i|^2} \; .
\label{eq:gamma}
\eeq
It is well-known that the SINR-maximizing receiver is the linear MMSE receiver, namely
\beq
\bd_k=\sqrt{p_k} h_k e^{j \varphi_k}\bM^{-1}\bs_k \; ,
\eeq
with $\bM= E\left\{ \br \br^H \right\}=
\ds \sum_{k=1}^K p_k h_k^2 \bs_k \bs_k^H + 2 \N \bI_N$ the covariance matrix of the data.

On the other hand, a WL detection structure can be also devised, i.e. the following detection rule can be considered:
\beq
\widehat{b}_k=\mbox{sign}\left\{\Re\left[\bd_{1,k}^H \br + \bd_{2,k}^H \br^*\right]\right\} \; ,
\label{eq:decrule2}
\eeq
with $(\cdot)^*$ denoting complex conjugate and where $\bd_{1,k}$ and $\bd_{2,k}$ are two $N$-dimensional vectors. Defining the augmented vectors
\begin{equation}
\br_a=\left[
\begin{array}{c}
\br\\
\br^*
\end{array}
\right]=\sum_{i=1}^K\sqrt{2p_i}h_i b_i\bs_{i,a}+\bn_a\\
\;, \quad \mbox{and} \; \, \, \bd_{k,a}=\left[
\begin{array}{c}
\bd_{1,k}\\
\bd_{2,k}
\end{array}
\right] \; ,
\end{equation}
with
\begin{equation}
\begin{split}
\bn_{a} &=\ds \left[
\begin{array}{c}
\bn\\
\bn^*
\end{array}
\right] \, \quad \mbox{and} \; \, \,
\bs_{{k,a}}=\ds \frac{1}{\sqrt2}\left[
\begin{array}{c}
\bs_ke^{j\varphi_k}\\
\bs^*_ke^{-j\varphi_k}
\end{array}
\right]
\end{split}
\end{equation}
the SINR corresponding to the decision rule (\ref{eq:decrule2}) is expressed as
\beq
\gamma_k=\ds \frac{2p_k h_k^2 |\bd_{k,a}^H \bs_{k,a}|^2}{2\N\|\bd_{k,a}\|^2 + \ds \sum_{i \neq k} 2p_i h_i^2
|\bd_{k,a}^H \bs_{i,a}|^2} \; ,
\label{eq:gamma2}
\eeq
The SINR maximizing receiver is now written as
\beq
\bd_{k,a}=\sqrt{2p_k}h_k{\bM_a}^{-1}\bs_{k,a}\; ,
\label{eq:WLMMSE}
\eeq
with $\bM_a=E\left\{\br_a \br_a^H\right\}=\ds \sum_{k=1}^K 2p_k h_k^2 \bs_{k,a} \bs_{k,a}^H + 2 \N \bI_{2N}$ the covariance matrix of the vector $\br_a$. It is easily shown that
\beq
\bM_a=\left[ \begin{array}{cc} \bM & \bM' \\ \bM'^* & \bM^* \end{array} \right] \; ,
\eeq
with $\bM'=\ds \sum_{k=1}^K p_k h_k^2  e^{2j\varphi_k} \bs_k \bs_k^T$ the pseudo-covariance matrix of the data vector $\br$.
The MSE achieved by the $k$-th user, is now defined as
\beq
{\rm MSE}_k  \triangleq E\left\{ |b_k - \bd_{k,a}^H \br_a|^2\right\} \; ,
\label{eq:mse}
\eeq
If receiver (\ref{eq:WLMMSE}) is used, it is easily shown that the $k$-th user's MSE is written as
\beq
{\rm MSE}_k  \triangleq E\left\{ |b_k - \bd_{k,a}^H \br_a|^2\right\} = \ds \frac{1}{1+\gamma_{k}} \; ,
\label{eq:rec}
\eeq
wherein $\gamma_k$ is the $k$-th user's SINR when receiver (\ref{eq:WLMMSE}) is used.

\section{Minimum-MSE non-cooperative joint spreading code and WL receiver  optimization}
Equipped with the above data model, in this section we consider the issue of non-cooperative joint spreading code and WL receiver optimization; the objective function here considered is the individual SINR and/or the individual MSE, since, as we will see, the problems of SINR-maximization and of MSE-minimization will turn out to be equivalent.

We thus start by considering the game
\beq
\ds \min_{\bs_k, \bd_{1,k}, \bd_{2,k}} {\rm MSE}_k \; , \quad k=1, \ldots, K-1 \; ,
\label{eq:gameWL}
\eeq
wherein $\bs_k$, $\bd_{1,k}$ and $\bd_{2,k}$ are $N$-dimensional complex vectors, with $\bs_k$ constrained to have unit norm.

To analyze this game, we first note that a joint minimization of the MSE with respect to the receive vectors and to the transmit spreading code cannot be done in closed form, so we resort to the usual iterative approach wherein MSE is minimized sequentially with respect to the receive vectors and to the spreading codes, and, after convergence has occurred, make an \emph{a posteriori} check that the global minimum of the cost function has been achieved.
We have already noted that the  WL receiver for the $k$-th user minimizing the $k$-th user's MSE in (\ref{eq:mse}) is given by the MMSE receiver in (\ref{eq:WLMMSE}).

Minimization of (\ref{eq:rec}) with respect to $\bs_{k,a}$, subject to the constraint that $\|\bs_{k,a}\|^2=1$, is straightforward, and gives the solution
\begin{equation}
\bs_{k,a}=\frac{\bd_{k,a}}{\|\bd_{k,a}\|} \; .
\end{equation}
Summing up, an iterative non-cooperative algorithm for MSE-minimization is the following:
\begin{equation}
\left\{
\begin{array}{lr}
\bd_{k,a}=\sqrt{2p_k}h_k{\bM_a}^{-1}\bs_{k,a}  & \quad \forall k=1,\ldots,K \, ,\\
\bs_{k,a}=\frac{\bd_{k,a}}{\|\bd_{k,a}\|}  &  \quad \forall k=1,\ldots,K \, .
\end{array} \right.
\label{eq:iterazionizz}
\end{equation}
Iteration (\ref{eq:iterazionizz}) resembles the well-known MMSE signature update algorithm
\begin{equation}
\left\{
\begin{array}{lr}
\bd_{k}=\sqrt{p_k}h_ke^{j \varphi_k }{\bM}^{-1}\bs_{k}  & \quad \forall k=1,\ldots,K \, ,\\
\bs_{k}=\frac{\bd_{k}}{\|\bd_{k}\|}  &  \quad \forall k=1,\ldots,K \, ,
\end{array} \right.
\label{eq:iterazioniul}
\end{equation}
which was first proposed in \cite{ulukusyates} and has been thoroughly analyzed in a number of subsequent studies such as \cite{rose,ulukusyener,ensuring}. As already noted, these results apply to the case of a real signal model with linear detection rule; iterations (\ref{eq:iterazionizz}), instead, come from the non-cooperative MSE minimization assuming a complex signal model and WL processing at the receiver. The sequel of this paper is thus devoted to the analysis of the properties of the steady-state solution of (\ref{eq:iterazionizz}) and to a performance analysis in comparison with the fixed point of iterations (\ref{eq:iterazioniul}).
Our algorithm starts with $K$ unit norm signatures $[\bs_{1,a}(0) \dots \bs_{K,a}(0)]$ that we collect into the matrix $\bS_a(0)$. At iteration $n$ the algorithm replaces each signature with its normalized MMSE filter, thus building the matrix $S_a(n)$.
Notice that each iteration is composed of $K$ intermediate steps. At the $k$-th intermediate step in iteration $n$ the first $k-1$ signatures have been updated to yield the matrix $S_{k-1,a}(n)=[\bs_{1,a}(n) \dots \bs_{(k-1),a}(n), \bs_{k,a}(n-1) \dots \bs_{K,a}(n-1)]$. Therefore the interference plus noise correlation matrix employed to update the signature $\bs_{k,a}$ is
\begin{equation}
\begin{array}{c}
\bA_k(n)=\sum_{j<k}2p_k{h_k}^2\bs_{j,a}(n)\bs_{j,a}^H(n)+
\sum_{j>k}2p_k{h_k}^2\bs_{j,a}(n-1)\bs_{j,a}^H(n-1)+2N_0\bI_{2N}\; .
\end{array}
\end{equation}
In the following we prove important theoretical results for the algorithm (\ref{eq:iterazionizz}).

\subsection{Proof of convergence of iterations (\ref{eq:iterazionizz})}
We start by showing that iterations (\ref{eq:iterazionizz}) always converge, i.e. that an NE exists for the considered game (\ref{eq:gameWL}). Recalling that a WL receiver operates according to the decision rule (\ref{eq:decrule2}),
and that its performance is ruled by the real part of the cross-correlation matrix of the adopted complex spreading codes, we
introduce the WL total weighted square Correlation (WL-TWSC) as follows:
\begin{equation}
\mbox{WL-TWSC} =\ds \sum_{i=1}^{K} \ds \sum_{j=1}^{K}p_i p_j h_i^2 h_j^2(\Re(\bs_i^H \bs_j))^2=\sum_{i=1}^{K}\sum_{j=1}^{K} p_i p_j h_i^2 h_j^2{|\bs_{i,a}^H \bs_{j,a}|}^2 \; .
\end{equation}
Since the WL-TWSC is lower bounded for any value of $K$,  the convergence of iterations (\ref{eq:iterazionizz}) can be easily proved by showing that they do not increase the WL-TWSC.
The WL-TWSC can be shown to be written as
\begin{equation}
\mbox{WL-TWSC}=p_k^2 h_k^4+2p_k h_k^2\bs_{k,a}^H  \bA_k \bs_{k,a}
-2p_k{h_k}^2{\sigma}^2+\beta_k
\end{equation}
where $\bA_k$ is the interference plus noise correlation matrix, and $\beta_k$ is the sum of the correlations not involving the signature $\bs_{k,a}$.
After replacing $\bs_{k,a}$ with $\bd_{k,a}/\|\bd_{k,a}\|=\frac{{\bA_k}^{-1}\bs_{k,a}}{({\bs_{k,a}^H  {\bA_k}^{-2} \bs_{k,a}})^{0.5}}$,
the WL-TWSC of the new set of signatures can be expressed as
\begin{equation}
\overline{\mbox{WL-TWSC}}=2p_k h_k^2\frac{{\bs_{k,a}}^H{\bA_k}^{-1}\bs_{k,a}}{
\bs_{k,a}^H  {\bA_k}^{-2} \bs_{k,a}}+\beta_k-2p_k h_k^2{\sigma}^2+p_k^2 h_k^4 \; .
\end{equation}
Next, we compute the EVD of $\bA_k=\bV \bLambda \bV^H$ and set $\bx=\bV^H \bs_{k,a}$.
Then,
\begin{equation}
\overline{\mbox{WL-TWSC}} \leq \mbox{WL-TWSC} \Longleftrightarrow
\bx^H \bLambda^{-1} \bx \leq (\bx^H \bLambda^{-2} \bx)(\bx^H \bLambda \bx) \; .
\end{equation}
Exploiting the results reported in the appendix of \cite{ulukusyates} it can be shown that
\begin{equation}
\begin{array}{c}
1 \leq (\bx^H \bLambda^{-1} \bx)(\bx^H \bLambda \bx) \; , \quad
\mbox{and} \quad (\bx^H \bLambda^{-1} \bx)^2 \leq (\bx^H \bLambda^{-2} \bx) \; .
\end{array}
\end{equation}
Applying these two bounds in succession it can be thus shown that the spreading code update decreases the WL-TWSC and, consequently, iterations (\ref{eq:iterazionizz}) admit a fixed point.

\subsection{Orthogonality of the signatures at the fixed point for $K\leq 2N$}
Generalizing the arguments of
\cite{ulukusyates} to the case in which complex spreading codes and WL MMSE filtering are adopted, we prove now that for $K\leq 2N$, the steady-state output of iterations (\ref{eq:iterazionizz}) is a set of orthonormal augmented signatures, provided that the $K$ augmented signatures used as a starting point of our algorithm are linearly independent. Of course, this condition can be met only if the range span of the $K$ starting (non-augmented) signatures $\bS(0)=[\bs_1, \ldots, \bs_K]$ coincides with ${\cal C}^N$, with $\cal C$ the complex field.

Let $\bA$ be the diagonal matrix whose $k$-th element on the diagonal is $2p_k h_k^2=a_k^2$, and
let $\bS=\left[\bs_{1,a}, \ldots, \bs_{K,a}\right]$.
We first prove the following lemma:
\begin{equation}
|x\bI_{2N}+\bar{\bS_a}\bA\bar{\bS_a}^{H}| \geq |x\bI_{2N}+\bS_a\bA\bS_a^H| \; , \quad \forall x \geq 0 \, ,
\label{eq:lemma}
\end{equation}
where $\bar{\bS_a}$ is the matrix of the augmented signatures after having updated $\bs_{k,a}$.
To prove (\ref{eq:lemma}), let us consider first the case $x>0$.
Set $\bZ_k=\sum_{j\neq k}a_j^2\bs_{j,a}\bs_{j,a}^H$; then
$$
\begin{array}{cc}
|x\bI_{2N}+\bS_a\bA\bS_a^H| =|x\bI_{2N}+\bZ_k+a_k^2\bs_{k,a}\bs_{k,a}^H|=
|x\bI_{2N}+\bZ_k| |\bI_{2N}+(x\bI_{2N}+\bZ_k)^{-1}a_k^2\bs_{k,a}\bs_{k,a}^H|=\\
|x\bI_{2N}+\bZ_k|(1+a_k^2\bs_{k,a}^H(x\bI_{2N}+\bZ_k)^{-1}\bs_{k,a}) \; .
\end{array}
$$
Similarly we obtain
\begin{equation}
\begin{array}{l}
|x\bI_{2N}+\bar{\bS_a}\bA\bar{\bS_a}^{H}|=
|x\bI_{2N}+\bZ_k|(1+a_k^2\frac{\bd_{k,a}^H}{\|\bd_{k,a}\|}(x\bI_{2N}+\bZ_k)^{-1}\frac{\bd_{k,a}}{\|\bd_{k,a}\|})
\end{array}
\end{equation}
Substituting for $\bd_{k,a}$ with its expression and taking into account the two previous equations,
lemma (\ref{eq:lemma}) can be expressed as
\begin{equation}
\begin{array}{llll}
\frac{\bs_{k,a}^H(\bZ_k+2N_0\bI_{2N})^{-1}(\bZ_k+x\bI_{2N})^{-1}
(\bZ_k+2N_0\bI_{2N})^{-1}\bs_{k,a}}{\bs_{k,a}^H(\bZ_k+2N_0\bI_{2N})^{-2}
\bs_{k,a}}
\geq \bs_{k,a}^H(x\bI_{2N}+\bZ_k)^{-1}\bs_{k,a}
\end{array}
\end{equation}
Upon performing the eigenvalue decomposition (EVD) of $\bZ_k=\bV\bLambda\bV^H$ and letting $\bw=\bV^H\bs_{k,a}$, we obtain
\begin{equation} \label{Lemma}
\sum_{i=1}^{2N}\frac{|w_i|^2}{(\lambda_i+x)(\lambda_i+2N_0)^2} \geq
\sum_{i=1}^{2N}\frac{|w_i|^2}{\lambda_i+x}\sum_{i=1}^{2N}\frac{|w_i|^2}{(\lambda_i+2N_0)^2}
\end{equation}
Since $\|\bw\|=1$, we are allowed to define a discrete random variable $Y$ with the following probability mass function:
\begin{equation}
p_Y(y)=\left\{
\begin{array}{lr}
|w_i|^2 & \qquad y=\lambda_i, 1 \leq i \leq N\\
0 & \qquad {\rm otherwise} \, .
\end{array} \right.
\end{equation}
Therefore, (\ref{Lemma}) is equivalently written as
\begin{equation} \label{Lemma3}
E\left[\frac{1}{(Y+x)(Y+2N_0)^2}\right] \geq E\left[\frac{1}{(Y+x)}\right] E\left[\frac{1}{(Y+2N_0)^2}\right] \; .
\end{equation}
For the proof of (\ref{Lemma3}) refer to \cite{ulukusyates}.
This proves the result for $x>0$. In order to prove the result for $x=0$, consider the following $N$th-order polynomial:
\begin{equation}
\begin{array}{lll}
f(x)=|x\bI_{2N}+\bar{\bS_a}\bA\bar{\bS_a}^{H}|-|x\bI_{2N}+\bS_a\bA\bS_a^H|=
\prod_{i=1}^{2N}(x+\bar{\lambda_i})-\prod_{i=1}^{2N}(x+\lambda_i)
\end{array}
\end{equation}
wherein $\bar{\lambda_i}$ and $\lambda_i$ are the eigenvalues of $\bar{\bS_a}\bA\bar{\bS_a}^{H}$ and $\bS_a\bA\bS_a^H$ respectively.
We have proved that $f(x) \geq 0, \forall x>0$. Then, due to the continuity of $f(x)$, we have $f(0) \geq 0$,
which proves the lemma for $x=0$.

Given the above lemma, it is straightforward to note that
\begin{equation}
|x\bI_{K}+\bA\bar{\bS_a}^{H}\bar{\bS_a}| \geq |x\bI_{K}+\bA\bS_a^H\bS_a|\; , \qquad \forall x \geq 0\, .
\end{equation}
Setting $x=0$ we obtain
\begin{equation}
|\bA\bar{\bS_a}^{H}\bar{\bS_a}| \geq |\bA\bS_a^H\bS_a| \; .
\end{equation}
Since $\bA$ is an invertible positive definite matrix, the rank of $\bA\bS_a^H\bS_a$ equals the rank of $\bS_a^H\bS_a$. Therefore,
if the initial matrix $\bS_a^H\bS_a$ has full rank, the matrix $\bS_a^H\bS_a$ at the fixed point will be invertible.

Let us now prove that at the fixed point the augmented signatures are eigenvectors of the correlation matrix
of the augmented received vector.
At iteration $n$ we can write
\begin{equation}
\bs_{k,a}(n)=\frac{\bM_a^{-1}(n-1)\bs_{k,a}(n-1)}{\|\bM_a^{-1}(n-1)\bs_{k,a}(n-1)\|} \; ,
\end{equation}
At the fixed point everything is independent of $n$. Therefore
\begin{equation}
\bs_{k,a}=\frac{\bM_a^{-1}\bs_{k,a}}{\|\bM_a^{-1}\bs_{k,a}\|}=\sigma_k\bM_a^{-1}\bs_{k,a} \; ,
\end{equation}
and finally we obtain
$\bM_a\bs_{k,a}=\sigma_k\bs_{k,a}$,
which proves the result.

Now, we exploit the above result in order to prove that at the fixed point we obtain an orthonormal set of signatures:
\begin{equation}
\bM_a\bs_{k,a}=(\bS_a\bA\bS_a^H+2N_0\bI_{2N})\bs_{k,a}=\sigma_k\bs_{k,a} \, ,\qquad \forall k \, .
\label{eq:fgfh}
\end{equation}
Multiplying both sides of (\ref{eq:fgfh}) by $\bs_{l,a}^H$ we have
\begin{equation} \label{orth}
\bs_{l,a}^H(\bS_a\bA\bS_a^H+2N_0\bI_{2N})\bs_{k,a}=\sigma_k\bs_ {l,a}^H\bs_{k,a} \, , \qquad \forall k,l \, .
\end{equation}
Rewriting (\ref{orth}) in a matrix form we obtain
\begin{equation}
\bS_a^H(\bS_a\bA\bS_a^H+2N_0\bI_{2N})\bS_a=\bS_a^H\bS_a\bSigma \; .
\end{equation}
After some algebra we have
\begin{equation}
\bS_a^H\bS_a=\bA^{-1}(\bSigma-2N_0\bI_{K}) \; .
\label{eq:aaaa}
\end{equation}
Since the matrix on the right-hand-side of (\ref{eq:aaaa}) is diagonal, at the fixed point we achieve orthogonal augmented signatures. Moreover, since the signatures are constrained to have unit norms, we conclude that the algorithm converges to an orthonormal augmented set. This is a very attractive feature, since transceiver optimization algorithms using linear receivers can provide an orthonormal set of signatures only for a number of users $K \leq N$, whereas, if WL processing can be used,  orthonormal signatures can be granted to a number of users $K \leq 2N$: this seemingly inexact results is due to the fact that, when WL processing is adopted what really matters is not the scalar product among the spreading codes, but its real part.

\subsubsection{Optimality of the steady-state solution for $k\leq 2N$}
Since the signatures are constrained to have unit norm, the following bound for the WL-TWSC holds:
\begin{equation} \label{Lower}
\mbox{WL-TWSC}\geq \frac{1}{4}\ds \sum_{k=1}^{K}a_k^4=\frac{1}{4}\rm{tr}(\bA^2) \; .
\end{equation}
It is easily verified that (\ref{Lower}) can be achieved when the signatures are orthonormal. Hence, when $K \leq 2N$ our algorithm achieves the global minimum of the WL-TWSC.
We also note that optimization of the WL-TWSC is equivalent to optimization of other performance measures such as the Total MMSE (TMMSE)\cite{ulukusyener}, and the Sum-Capacity ($C_{\rm sum}$)\cite{Csum}.
Indeed, the WL-TWSC, TMMSE and $C_{\rm sum}$ for the considered system can be easily expressed in terms of the eigenvalues of $\bS_a\bA\bS_a^H$, $[\lambda_1 \ldots \lambda_{2N}]$ as follows:
\begin{equation}
\begin{array}{l}
\mbox{WL-TWSC}=\frac{1}{4}\ds\sum_{i=1}^{2N}\lambda_i^2 \; ,\\
C_{\rm sum}=\ds\sum_{i=1}^{2N}\log(1+\lambda_i/2N_0) \; \; \; \mbox{and}\\
{\rm TMMSE}=\ds K-\sum_{i=1}^{2N}\frac{\lambda_i}{\lambda_i+2N_0} \; .
\end{array}
\end{equation}
Since $\bS_a\bA\bS_a^H$ is an hermitian matrix, its eigenvalues will be real numbers. As a consequence, the WL-TWSC and TMMSE are Schur-convex functions, while $C_{\rm sum}$ is a Schur-concave one. Therefore, we conclude that the signature set provided by the fixed point of (\ref{eq:iterazionizz})  also minimizes the TMMSE and maximizes $C_{\rm sum}$.

\subsection{Properties of the steady-state spreading codes for $K>2N$}
So far, we have proved that the algorithm (\ref{eq:iterazionizz}) always converges, and that it achieves the global minimum of the WL-TWSC for $K\leq2N$. Now, properties of the equilibrium spreading-codes for $K>2N$ are investigated.

We assume that the algorithm is initialized with a full rank matrix $\bS_a$, so that at the fixed point the matrix
$\bS_a\bA\bS_a^H$ is positive definite.
We start with the following theorem that characterizes the fixed points of (\ref{eq:iterazionizz}).
\begin{theorem}
Let $\bS_a$ be a fixed point of (\ref{eq:iterazionizz}), and {$\bq_1,\ldots,\bq_N$} be an orthonormal basis of eigenvectors of $\bS_a\bA\bS_a^H$. Then, there exist a number $L\in\{1,\ldots,N\}$, a partition $\{J_1,\ldots,J_L\}$ of the set $\{1,\ldots,K\}$, a partition $\{I_1,\ldots,I_L\}$ of the set $\{1,\dots,N\}$, and positive numbers $\lambda_1,\ldots,\lambda_L$, such that
\begin{equation}
\lambda(\bS_a\bA\bS_a^H)=[\underbrace{\lambda_1,\ldots,\lambda_1}_{|I_1|},\ldots,
\underbrace{\lambda_L,\ldots,\lambda_L}_{|I_L|}] \; ,\label{eigenvalue}
\end{equation}
\begin{equation}
\lambda_{\ell}=\frac{1}{|I_{\ell}|}\sum_{k\in J_{\ell}}a_k^2 \; , \qquad \forall \ell \in {1,\ldots,L} \; ,
\end{equation}
\begin{equation}
\bs_{k,a_1}^H\bs_{k,a_2}=0 \; , \qquad \forall k_1 \in J_{\ell}, k_2 \notin J_{\ell} \quad \mbox{and} \label{herm}
\end{equation}
\begin{equation}
\{\bs_{k,a}:k \in J_{\ell}\}\subset {\rm span}\{\bq_n: n \in I_{\ell}\} \; ,
\end{equation}
where $|I_l|$ denotes the cardinality of $I_l$, and the sets $J_l$ and $I_l$ are defined as follows:
\begin{equation}
J_l=\{k\in{1,\ldots,K}:\bS_a\bA\bS_a^H\bs_{k,a}=\lambda_l\bs_{k,a}\}  \quad \mbox{and} \label{PartK}
\eeq
\beq
I_l=\{n\in{1,\ldots,N}:\bS_a\bA\bS_a^H\bq_n=\lambda_l\bq_n\}\label{PartN} \, .
\end{equation}
\end{theorem}
\proof
A complete proof for the case of real channels can be found in \cite{ensuring}. Here we extend it to the complex case, putting emphasis on some points that will be useful in the sequel.
Let $L$ be the number of distinct eigenvalues of $\bS_a\bA\bS_a^H$. From previous sections we know that all of the augmented vectors
$\bs_{k,a}$ are eigenvectors of $\bS_a\bA\bS_a^H$.
By grouping the augmented signatures associated to the same eigenvalues we obtain (\ref{PartK}).
Since $\bS_a\bA\bS_a^H$ is an hermitian matrix, eigenvectors associated with distinct eigenvalues are orthogonal, thus implying (\ref{herm}).
Now, let ${\bq_1,\ldots,\bq_N}$ be an orthonormal basis of eigenvectors of the hermitian positive definite matrix $\bS_a\bA\bS_a^H$. If we group together the eigenvectors $\bq_n$ associated with the same eigenvalue, we obtain (\ref{PartN}).
The cardinality of  $I_{\ell}$ equals the  (geometric and algebraic) multiplicity of the eigenvalue $\lambda_{\ell}$. Then, (\ref{eigenvalue}) is proved.
The interested reader can refer to \cite{ensuring} for the rest of the proof.
\endproof

Summing up, this theorem states that at the fixed point, the matrix $\bS_a\bA\bS_a^H$ has $L$ distinct eigenvalues with multiplicities $|I_1|,\dots,|I_L|$ respectively, and the columns of the matrix $\bS_a$ are partitioned into orthogonal sets, with each set containing the columns that are eigenvectors associated with the same eigenvalue of $\bS_a\bA\bS_a^H$.
Note that since we are assuming a full-rank matrix $\bS_a$, and $K>2N$, the columns of $\bS_a$ will span $\mathcal{C}^{2N}$, implying that for each eigenvalue $\lambda_{\ell}$ there will be at least a column of $\bS_a$ that is an eigenvector associated with that eigenvalue. This means that we can always choose the orthonormal basis ${\bq_1,\ldots,\bq_N}$ so that each $I_{\ell}$ contains at least one proper augmented vector.
Moreover, since $K>2N$, and $\bS_a\bA\bS_a^H$ is not singular, denoting by $\bS_{J_{\ell}}$ the matrix whose columns are the vectors belonging to $J_{\ell}$, we have $|I_{\ell}|={\rm rank}(\bS_{J_{\ell}})$.

\subsubsection{The suboptimal fixed points of iterations (\ref{eq:iterazionizz}) are unstable}
Reference \cite{ensuring} proves that for the case of real signals and linear processing all the suboptimal fixed points of the MMSE update algorithm are unstable.
We  prove here that a similar result also holds true for iterations (\ref{eq:iterazionizz}).
As (\ref{eq:iterazionizz}) does not increase the WL-TWSC, all we need
to prove is that among the fixed points of (\ref{eq:iterazionizz}) there are no local minima of the WL-TWSC, except the global minimum.

To begin with, let $\mathcal{V}$ and $\mathcal{S}$ denote the sets $\{\bv_a: \bv_a=\left[
\begin{array}{c}
\bv\\
\bv^*
\end{array}\right],
\|\bv_a\|=1, \bv \in \mathcal{C}^N\},$ and
$\{[\bs_{1,a},\ldots,\bs_{K,a}]: \bs_{k,a} \in \mathcal{V}\}$, respectively.
We can define a metric on $\mathcal{S}$ as
$$d:(\bS'_a,\bS''_a) \in \mathcal{S}^2 \rightarrow d(\bS'_a,\bS''_a)=\max_{k=1,\ldots,K} \arccos({\bs'}_{k,a}^H\bs''_{k,a}) \, .$$
Note that for vectors in $\mathcal{C}^{2N}$, ${\bs'}_{k,a}^H\bs''_{k,a}$ is in general a complex quantity, and hence $d$ is not a metric.
However, we now prove that as long as vectors in $\mathcal{V}$ are considered, ${\bs'}_{k,a}^H\bs''_{k,a}$ is real and $d$ is a metric on $\mathcal{S}$.

Let $\bv_{1,a}=\left[
\begin{array}{c}
\bv_1\\
\bv_1^*
\end{array}\right]$
and $\bv_{2,a}=\left[
\begin{array}{c}
\bv_2\\
\bv_2^*
\end{array}\right]$
be two generic vectors belonging to $\mathcal{V}$.
Then we have
\begin{equation}
{\bv}_{1,a}^H\bv_{2,a}=2\Re{(\bv_1^H\bv_2)}
\end{equation}
which is of course real.
Now, if we express $\bv_1$ and $\bv_2$ in terms of their real and imaginary parts, we obtain
\begin{equation}
\begin{array}{llll}
{\bv}_{1,a}^H\bv_{2,a} =
 2\Re{(\bv_1^H\bv_2)}   =2(\bv_{R,1}^T\bv_{R,2}+\bv_{I,1}^T\bv_{I,2})=
\bv_{RI,1}^T\bv_{RI,2}
\end{array}
\end{equation}
where
$\bv_{R,1}$, and $\bv_{I,1}$ are the real and imaginary part of the vector $\bv_1$;
$\bv_{R,2}$, and $\bv_{I,2}$ are the real and imaginary part of the vector $\bv_2$;
$\bv_{RI,1}=\sqrt2\left[
\begin{array}{c}
\bv_{R,1}\\
\bv_{I,1}
\end{array}\right]$,
and $\bv_{RI,2}=\sqrt2\left[
\begin{array}{c}
\bv_{R,2}\\
\bv_{I,2}
\end{array}\right]$.
Therefore, when vectors in $\mathcal{V}$ are considered, we can always express the hermitian product $\bv_{1,a}^H\bv_{2,a}$ in terms of a scalar product between real vectors. Moreover, since
$\bv_{1,a}$ and $\bv_{2,a}$ have unit norm, the vectors $\bv_{RI,1}$ and $\bv_{RI,2}$ will also have unit norm.
As a consequence we have
\begin{equation}
\begin{array}{l}
\arccos(\bv^H_{1,a} \bv_{2,a})=\arccos(\bv_{RI,1}^T\bv_{RI,2})=
\arccos(\|\bv_{RI,1}\|\|\bv_{RI,2}\|\cos(\alpha))=\alpha \in[0,\pi],
\end{array}
\end{equation}
where $\alpha$ is the angle between the vectors $\bv_{RI,1}$ and $\bv_{RI,2}$.
This allows us to prove that $d$ satisfies the triangle inequality.
In fact, given $\bS_a,\bS'_a,\bS''_a \in \mathcal{S}$ we have
\begin{equation}
\begin{array}{cc}
d(\bS_a,\bS''_a)=\max_{k=1,\ldots,K} \arccos({\bs}_{k,a}^H\bs''_{k,a})=
\ds\max_{k=1,\ldots,K}\arccos({\bs}_{RI,k}^T\bs''_{RI,k})\leq\\
\ds\max_{k=1,\ldots,K}\left[\arccos({\bs}_{RI,k}^T\bs'_{RI,k})+\arccos({\bs'}_{RI,k}^T\bs''_{RI,k})\right]\leq \\
\ds \max_{k=1,\ldots,K}\arccos({\bs}_{k,a}^T\bs'_{k,a})+\max_{k=1,\ldots,K}\arccos({\bs'}_{k,a}^T\bs''_{k,a})
=d(\bS_a,\bS'_a)+d(\bS'_a,\bS''_a) \; .
\end{array}
\end{equation}
The other properties that characterize a metric are straightforward to prove, and hence $d$ is a metric on $\mathcal{S}$.\\
In the following we extend to the case at hand some results proved in \cite{ensuring} for real signals.
\begin{lemma}
Let $\bS_a\in\mathcal{S}$ be a fixed point of (\ref{eq:iterazionizz}) and a local minimum of WL-TWSC.
Then, given $\ell_1$,$\ell_2\in{1,\ldots,L}$ with
$\lambda_{\ell_1}>\lambda_{\ell_2}$, $k_1\in J_{\ell_1}$, $k_2\in J_{\ell_2}$ we must have $a_{k_1}^2\geq a_{k_2}^2$.
\end{lemma}
\proof
Suppose the thesis does not hold. Then, take $\epsilon>0$ and set $\alpha=\sin(\epsilon)$,
$\beta=\frac{-a_{k_1}^2}{a_{k_2}^2}\alpha$.
Now consider a matrix $\bS'_a$ with the same columns of $\bS_a$ except for
$$\bs'_{k_1,a}=\sqrt{1-\alpha^2}\bs_{k_1,a}+\alpha\bs_{k_2,a} \;  \quad
\mbox{and} \quad
\bs'_{k_2,a}=\sqrt{1-\beta^2}\bs_{k_2,a}+\beta\bs_{k_1,a}\; .$$
Note that $\bS'_a\in\mathcal{S}$ because $\bs_{k_1,a}$ and $\bs_{k_2,a}$ are orthogonal, which implies $\|\bs'_{k_1,a}\|=\|\bs'_{k_2,a}\|=1$. Moreover, $\alpha$, $\beta\in\mathcal{R}$.
Therefore, $\bs_{k_1,a}$ and $\bs_{k_2,a}$ are proper augmented vectors.
Since we are assuming $a_{k_1}^2 < a_{k_2}^2$, it is easily seen that
$d(\bS'_a,\bS_a)\leq\epsilon$, which means $\bS'_a$ belongs to the $\epsilon$-neighborhood of $\bS_a$.
Now we set
$\bS'_a\bA\bS^{'H}_a =\bS_a\bA\bS^{H}_a +\bDelta$\\
where
\begin{equation}
\begin{array}{lll}
\bDelta=&(\beta^2 a_{k_2}^2-\alpha^2 a_{k_1}^2)(\bs_{k_1,a}\bs_{k_1,a}^H-\bs_{k_2,a}\bs_{k_2,a}^H ) + \left(a_{k_1}^2\alpha\sqrt{1-\alpha^2}+a_{k_2}^2\beta\sqrt{1-\beta^2}\right)
 + (\bs_{k_1,a}\bs_{k_2,a}^H+\bs_{k_2,a}\bs_{k_1,a}^H)\; .
\end{array}
\end{equation}
Then, after some manipulation we get
$$
\begin{array}{lll}
\ds \mbox{WL-TWSC}(\bS'_a)-\mbox{WL-TWSC}(\bS_a)= \frac{1}{4} \left(\ds 2{\rm tr}(\bS_a\bA\bS_a^H\bDelta)+{\rm tr}(\bDelta^2)=  \ds 2(\lambda_{\ell_1}-\lambda_{\ell_2})\epsilon^2 a_{k_1}^2\left(1-\frac{a_{k_1}^2}{a_{k_2}^2}\right)+o(\epsilon^3)\right) \; .
\end{array}
$$
Given the hypothesis, and as we are assuming $a_{k_1}^2 < a_{k_2}^2$, for $\epsilon$ approaching $0$
we have $\mbox{WL-TWSC}(\bS_a)>\mbox{WL-TWSC}(\bS'_a)$.
Therefore, we have found a matrix in the $\epsilon$-neighborhood of $\bS_a$, with a higher WL-TWSC
than $\bS_a$, which contradicts $\bS_a$ being a local minimum of WL-TWSC.
\endproof

\begin{lemma}
Let $\bS_a\in\mathcal{S}$ be a fixed point of (\ref{eq:iterazionizz}) and a local minimum of WL-TWSC.
Let $\ell \in {1,\ldots,L}$ with
$\lambda_{\ell}>\min_{{\ell}' \in {1,\ldots,L}}\lambda_{{\ell}'}$. Then $|J_{\ell}|\leq |I_{\ell}|$.
\end{lemma}

\proof
Suppose the thesis does not hold. Then, we can find $\ell_1$, $\ell_2\in{\,\ldots,L}$, with
$\lambda_{\ell_1}>\lambda_{\ell_2}$ and $|J_{\ell_1}|> |I_{\ell_2}|$.
As a consequence, since $|J_{\ell_1}|>|I_{\ell_1}|=\rm{rank}(\bS_{J_{\ell_1}})$,
there exists a unit norm vector $\bv\in\mathcal{C}^{|J_{\ell_1}|}$, belonging to the null space of $\bS_{J_{\ell_1}}\bA_{J_{\ell_1}}$.
We now show that, for the case at hand, the null space of $\bS_{J_{\ell_1}}\bA_{J_{\ell_1}}$ also contains a unit norm vector $\bv_r\in\mathcal{R}^{|J_{\ell_1}|}$.
Indeed, given the structure of the matrix $\bS_{J_{\ell_1}}\bA_{J_{\ell_1}}$, and letting $\mathcal{N}$ be the null space of $\bP=\bS_{J_{\ell_1}}\bA_{J_{\ell_1}}$, we can write
\begin{equation}
\bP=\left[\left(
\begin{array}{c}
\bp_1\\
\bp_1^*
\end{array}\right),\ldots,
\left(
\begin{array}{c}
\bp_{|J_{\ell_1}|}\\
\bp_{|J_{\ell_1}|}^*
\end{array}\right)
\right]\; .
\end{equation}
Then,
\begin{equation}
\bv\in\mathcal{N} \Leftrightarrow
\left\{
\begin{array}{l}
\ds\sum_{i=1}^{|J_{\ell_1}|}v_i\bp_i=0\\
\ds\sum_{i=1}^{|J_{\ell_1}|}v_i\bp_i^*=0 \; .
\end{array}
\right.
\end{equation}
Conjugating the second equation, and summing and subtracting, we obtain
\begin{equation}
\bv\in\mathcal{N} \Leftrightarrow
\left\{
\begin{array}{l}
\ds\sum_{i=1}^{|J_{\ell_1}|}\Re{(v_i)}\bp_i=0\\
\ds \sum_{i=1}^{|J_{\ell_1}|}\Im{(v_i)}\bp_i^*=0 \; .
\end{array}
\right.
\end{equation}
Therefore, the vector $\bv_r=\Re{(\bv)}/\|\Re{(\bv)}\|$, is a real unit norm vector belonging to $\mathcal{N}$, whose components we denote by $[v_{r,1},\ldots,v_{r,k},\ldots,v_{r,{|J_{\ell_1}|}}]$.
Now take $\epsilon>0$ and define a matrix $\bS'_a$ with the same columns of $\bS_a$ for $k\notin J_{\ell_1}$, and $\bs'_{k,a}=(\cos{\alpha_k})\bs_{k,a}+(\sin{\alpha_k})\bq$, for $k\in J_{\ell_1}$, where $\alpha_k=\epsilon v_{r,k}$,
and $\bq\in \mathcal{V}\bigcap I_{\ell_2}$. As pointed out in the previous section, such a vector always exists.
Therefore, $\bs'_{k,a}\in \mathcal{V}$, which implies $\bS'_a\in\mathcal{S}$.
Note also that $d(\bS_a,\bS'_a)=\epsilon\max_{k\in J_{\ell_1}}|v_{r,k}|\leq\epsilon$, implying that $\bS'_a$ is in the $\epsilon$-neighborhood of $\bS_a$.
Again, perturbing the signature matrix, we obtain
$\mbox{WL-TWSC}(\bS_a)-\mbox{WL-TWSC}(\bS'_a)=\frac{1}{4}\left(
2\epsilon^2(\lambda_{\ell_1}-\lambda_{\ell_2})\|\bA_{J_{\ell_1}}\bv_r\|^2+o(\epsilon^3)\right)$.
Since for $\epsilon$ approaching $0$ we have $\mbox{WL-TWSC}(\bS_a)>\mbox{WL-TWSC}(\bS'_a)$, we can find a matrix in the $\epsilon$-neighborhood of $\bS_a$ with a larger WL-TWSC, which contradict $\bS_a$ being a local minimum of WL-TWSC.
\endproof

Similarly to \cite{ensuring}, these results can be exploited to prove that among the fixed points of (\ref{eq:iterazionizz}) there are no local minima of the WL-TWSC except the global minimum, implying that all the non-optimal fixed points of our algorithm are unstable. Hence, a noisy version of (\ref{eq:iterazionizz}), obtained by adding a small perturbation to the deterministic updates, will converge to the global minimum of the WL-TWSC with probability 1.

\subsection{Oversized users and WL filtering}
Reference \cite{optimal} shows that, for the case in which $K>N$ and real signatures with linear detection are considered, the optimal spreading code allocation may grant to some large-power users orthogonal channels. Otherwise stated, if one or more users are received with a power that is large relative to that of the remaining ones, these users, called \emph{oversized users}, are given orthogonal spreading codes and enjoy single-user performance; the remaining users, instead are given generalized Welch-Bound-Equality (WBE) sequences and are confined in the orthogonal complement of the subspace spanned by these oversized users. More precisely, in \cite{optimal}, user $i$ is defined to be oversized if
\begin{equation}\label{Oversized}
d_i^2>\frac{\sum_{j=1}^K d_j^2 1_{d_i^2>d_j^2}}{N-\sum_{j=1}^K 1_{d_j^2\geq d_i^2}}
\end{equation}
where $d_i^2=p_ih_i^2 \quad \forall i=1,\ldots,K$, and $1_{d_n^2\geq d_m^2}=1$ if $d_n^2\geq d_m^2$ and zero otherwise.
Obviously, there can be at most $N-1$ oversized users.

Now, let us focus on the WL filtering case with complex spreading codes. We have seen that for $K\leq 2N$ it is possible to grant each user a spreading code that is mapped into an orthogonal augmented signature at the receiver, which is of course the optimum spreading code allocation. As a consequence, for $N<K\leq 2N$ the distinction between oversized and non-oversized users is meaningless if WL filtering is used, while it is necessary in the linear filtering case.
Consider now the more interesting case in which $K>2N$.
When WL filtering is considered, it can be shown that  a user is oversized if
\begin{equation}
a_i^2>\frac{\sum_{j=1}^K a_j^2 1_{a_i^2>a_j^2}}{2N-\sum_{j=1}^K 1_{a_j^2\geq a_i^2}} \; .
\label{eq:ovWL}
\end{equation}
Note that now there can be at most $2N-1$ oversized users.
Again, the optimum spreading code allocation is to grant each oversized user an orthogonal augmented spreading code, while other users are given generalized WBE sequences.

Now, comparing relations (\ref{Oversized}) and (\ref{eq:ovWL}) it can be shown that the set of oversized users resulting from iterations (\ref{eq:iterazionizz}) includes the set of oversized users resulting from the use of iterations (\ref{eq:iterazioniul}).
Indeed, assuming user $i$ satisfies (\ref{Oversized}), we can write
\begin{equation}
\begin{split}
d_i^2 &>\frac{\sum_{j=1}^K d_j^2 1_{d_i^2>d_j^2}}{N-\sum_{j=1}^K 1_{d_j^2\geq d_i^2}}\Rightarrow
d_i^2 >\frac{\sum_{j=1}^K d_j^2 1_{d_i^2>d_j^2}}{2N-\sum_{j=1}^K 1_{d_j^2\geq d_i^2}}\Leftrightarrow\\
\Leftrightarrow 2d_i^2 &>\frac{\sum_{j=1}^K 2d_j^2 1_{d_i^2>d_j^2}}{2N-\sum_{j=1}^K 1_{d_j^2\geq d_i^2}}\Leftrightarrow
a_i^2 >\frac{\sum_{j=1}^K a_j^2 1_{a_i^2>a_j^2}}{2N-\sum_{j=1}^K 1_{a_j^2\geq a_i^2}} \; .
\end{split}
\end{equation}
Accordingly, denoting by $\cal{K}_{\rm L}$ and $\cal{K}_{\rm WL}$ the sets of oversized users in the linear and WL case respectively, we have the relation $\cal{K}_{\rm L}\subseteqq \cal{K}_{\rm WL}$. Of course, having more oversized users is a good occurrence, since more users will have orthogonal channels.

\subsection{Equivalence between the minimum-MSE and the maximum-SINR non-cooperative games}
Equation (\ref{eq:rec}) shows the link between the $k$-th user's MSE and its achieved SINR under the assumption that the WL MMSE receiver is used. Given this one-to-one strictly decreasing relationship, it thus follows that non-cooperative MSE minimization is equivalent to non-cooperative SINR maximization.
A different, yet equivalent form for the game (\ref{eq:gameWL}) is thus
\beq
\ds \max_{\bs_k, \bd_{1,k}, \bd_{2,k}} \gamma_k \; , \quad k=1, \ldots, K-1 \; .
\label{eq:game2WL}
\eeq

\subsection{Sum-capacity of a CDMA system using as spreading codes the fixed point of iterations (\ref{eq:iterazionizz})}
The sum-capacity of a CDMA system with processing gain $N$ using the spreading codes resulting from the unique stable fixed point of iterations (\ref{eq:iterazionizz}) can be contrasted to that of a CDMA system using complex $2N$-dimensional codes with linear detection.
In this section we thus compare the sum-capacity of the following two systems:
\begin{itemize}
\item [a)] a CDMA system with processing gain $N$ and a WL receiver. This system can be seen as a CDMA system employing $2N$-dimensional, unit-norm, augmented spreading codes, that is codes in $\mathcal{S}$.
\item [b)] a CDMA system employing $2N$-dimensional, unit-norm, complex spreading codes and linear detection.
\end{itemize}
For the sake of comparison, we suppose that the noise power, the channel gains, and transmit powers are the same. Therefore, the parameter $\sigma$, and the matrix $\bA$ are the same for both systems.
Now, denote by $\bS_c$ and $\mathcal{C}_1^{2N \times K}$ the spreading matrix of system b), and the set of all matrixes in $\mathcal{C}^{2N \times K}$ with unit-norm columns, respectively. The sum-capacities of the two systems can be written as \cite{rupf}
\begin{equation}
\begin{split}
C_{{\rm sum,WL}}&=\max_{\bS_a\in\mathcal{S}}{\rm log}|\bI_{2N}+\sigma^{-2}\bS_a\bA\bS_a^H| \quad \mbox{and}\\
C_{\rm sum,C}&=\max_{\bS_c\in\mathcal{C}_1^{2N \times K}}{\rm log}|\bI_{2N}+\sigma^{-2}\bS_c\bA\bS_c^H| \; .
\end{split}
\end{equation}
Since $\mathcal{S}\subset\mathcal{C}_1^{2N \times K}$, it is clear that $C_{\rm sum,WL} \leq C_{\rm sum,C}$. However, we will now prove that $C_{\rm sum,WL}=C_{\rm sum,C}$, and therefore no loss in performance (in terms of sum-capacity) is suffered when using codes in $\mathcal{S}$.
To see this, we will consider separately the cases $K\leq2N$ and $K>2N$.
Let us first assume that $K\leq2N$. We can write
\begin{equation}
\begin{split}
C_{\rm sum,C}&=\max_{\bS_c\in\mathcal{C}_1^{2N \times K}}{\rm log}|\bI_{2N}+\sigma^{-2}\bS_c\bA\bS_c^H|=\max_{\bS_c\in\mathcal{C}_1^{2N \times K}}{\rm log}|\bI_{K}+\sigma^{-2}\bS_c^H\bS_c\bA|={\rm log}|\bI_K+\sigma^{-2}\bA|
\end{split}
\end{equation}
In the last equality we have exploited the result that for $K\leq2N$ the maximum is achieved by orthonormal signatures.
As for $C_{\rm sum,WL}$ we have
\begin{equation}
C_{\rm sum,WL}=\max_{\bS_a\in\mathcal{S}}{\rm log}|\bI_{2N}+\sigma^{-2}\bS_a\bA\bS_a^H|=
\max_{\bS_a\in\mathcal{S}}{\rm log}|\bI_{K}+\sigma^{-2}\bS_a^H\bS_a\bA| \; .
\end{equation}
Again, the maximum is achieved by a matrix with orthonormal columns. From previous sections we know that such a matrix is always to be found in $\mathcal{S}$ for $K\leq2N$, and therefore we can write
\begin{equation}
C_{\rm sum,WL}=\max_{\bS_a\in\mathcal{S}}{\rm log}|\bI_{K}+\sigma^{-2}\bA|=C_{\rm sum,C}\; .
\end{equation}

Consider now the case $K>2N$. First, we express the sum-capacity of both systems in terms of the vectors $\blambda(\bS_c\bA\bS_c^H)$ and $\blambda(\bS_a\bA\bS_a^H)$, whose components are the eigenvalues of the matrix $\bS_c\bA\bS_c^H$ and $\bS_a\bA\bS_a^H$, respectively:
\begin{equation}
C_{\rm sum,C}=\max_{\blambda(\bS_c\bA\bS_c^H)}\sum_{i=1}^{2N}{\rm log}(1+(\lambda_i/\sigma^2)) \quad \mbox{and}
\end{equation}
\begin{equation}
C_{\rm sum,WL}=\max_{\blambda(\bS_a\bA\bS_a^H)}\sum_{i=1}^{2N}{\rm log}(1+(\lambda_i/\sigma^2))\; .
\end{equation}
In order to proceed with the proof, we need a preliminary result first. We need to prove that $C_{\rm sum,WL}=2C_{\rm sum,R}$, where $C_{\rm sum,R}$ is the sum-capacity of a CDMA system employing $2N$-dimensional, real, unit-norm, spreading codes.
Denoting by $\bS_r$ and $\mathcal{R}_1^{2N \times K}$ the real-valued spreading matrix, and the set of all matrixes in $\mathcal{R}^{2N \times K}$ with unit-norm columns, respectively, we have
\begin{equation}
C_{\rm sum,R}=\max_{\bS_r\in \mathcal{R}_1^{2N \times K}} \frac{1}{2}{\rm log}|\bI_{2N}+\sigma^{-2}\bS_r\bA\bS_r^T|=\max_{\bS_r\in \mathcal{R}_1^{2N \times K}} \frac{1}{2}{\rm log}|\bI_{K}+\sigma^{-2}\bS_r^T\bS_r\bA| \; .
\end{equation}
Note that \cite{ulukusyener}, for real spreading matrixes the sum-capacity is defined with a factor $1/2$.

Now, consider the following bi-injective transformation between the sets $\mathcal{S}$ and $\mathcal{R}_1^{2N \times K}$
\begin{equation}\label{Transform}
f:\bS_a=\left[
\begin{array}{ll}
\bv_1\ldots\bv_K\\
\bv_1^*\ldots\bv_K^*
\end{array}\right]
\in\mathcal{S}\rightarrow\bS_r=
\left[
\begin{array}{ll}
\Re(\bv_1)\ldots\Re(\bv_K)\\
\Im(\bv_1)\ldots\Im(\bv_K)
\end{array}\right] \in\mathcal{R}_1^{2N \times K} \; .
\end{equation}
This implies that for any matrix $\bS_a \in \mathcal{S}$, there exists a matrix $ \bS_r=f(\bS_a) \in \mathcal{R}_1^{2N \times K}$ such that
\begin{equation} \label{Product}
\bS_a^H\bS_a=\bS_r^T\bS_r \; .
\end{equation}
Therefore we can write
\begin{equation}\label{Csum_WL_R}
C_{\rm sum,WL}=\max_{\bS_a\in \mathcal{S}}{\rm log}|\bI_{K}+\sigma^{-2}\bS_a^H\bS_a\bA|
=\max_{\bS_r\in \mathcal{R}_1^{2N \times K}} {\rm log}|\bI_{K}+\sigma^{-2}\bS_r^T\bS_r\bA|=2C_{\rm sum,R} \, .
\end{equation}
Bearing this result in mind we can resume our main proof. In fact, generalizing the arguments of \cite{optimal} to the case of complex $2N$-dimensional spreading codes, it can be shown that both the eigenvalues $\blambda(\bS_c\bA\bS_c^H)$ as $\bS_c$ takes value in $\mathcal{C}_1^{2N \times K}$, and the eigenvalues $\blambda(\bS_r\bA\bS_r^T)$ as $\bS_r$ takes value in $\mathcal{R}_1^{2N \times K}$, define the set
$$\{(\lambda_1,\ldots,\lambda_{2N})\in \mathcal{R}_+^{2N}:(\lambda_1,\ldots,\lambda_{2N},0,\ldots,0) \;\,{\rm majorizes} \;\, (a_1,\ldots,a_K)\}\; ,$$
which implies $C_{\rm sum,C}=2C_{\rm sum,R}$. Thus, by virtue of (\ref{Csum_WL_R}), we get $C_{\rm sum,C}=2C_{\rm sum,R}=C_{\rm sum,WL}$.

Therefore, a CDMA system with processing gain $N$, a WL receiver and spreading codes resulting from iterations (\ref{eq:iterazionizz}), has the same sum-capacity of a CDMA system with processing gain $2N$, a linear receiver and spreading codes resulting from iterations (\ref{eq:iterazioniul}).

\section{Non-cooperative power and transceiver optimization for maximum energy-efficiency}
So far, we have considered maximum-SINR and minimum-MSE games with respect to the spreading code and WL receiver of every user. Now, we include the transmit power among the parameters to be tuned, and consider the non-cooperative power control and WL transceiver design game assuming that the objective function is the energy efficiency, i.e. the number of bits reliably delivered to the receiver for each energy unit used for transmission.

To be more definite, assume that each mobile terminal sends its data in packets of $M$ bits, and that it is interested both in having its data received with as small as possible error probability at the AP, and in making  careful use of the energy stored in its battery. Obviously, these are conflicting goals, since error-free reception may be achieved by  increasing the transmit power, which of course comes at the expense of battery life. A useful approach to quantify these conflicting goals is to define the utility of the $k$-th user as the ratio of its throughput, defined as the number of information bits that are received with no error in unit time, to its transmit power \cite{meshkati,buzzijsac}, i.e.
\beq
u_k={T_k}/{p_k}\; .
\label{eq:utility}
\eeq
Note that $u_k$ is measured in bits/Joule, i.e. it represents the number of successful bit transmissions that can be made for each battery energy-unit used for transmission.
The utility function (\ref{eq:utility}) is widely accepted and indeed it has been already used in a number of previous studies such as \cite{nara2,meshkati,bacci,buzzijsac}.
Denoting by $R$ the common rate of the network  and assuming that each packet of $M$ symbols contains $L$ information symbols and $M-L$ overhead symbols, reserved, e.g., for channel estimation and/or parity checks,  and following the reasoning of
\cite{nara2,meshkati}, a faithful and mathematically tractable approximation for the utility $u_k$ in (\ref{eq:utility}) is the following:
\beq
u_k=R \ds \frac{L}{M} \frac{f(\gamma_k)}{p_k} \; , \quad \forall k=1, \ldots, K \; .
\label{eq:utility2}
\eeq
In the above equation, $f(\gamma_k)$ is the so-called \emph{efficiency function}, approximating the probability of successful (i.e. error-free) packet reception. As an example, for binary phase-shift keying (BPSK) modulation, the choice $f(\gamma_k)=(1-e^{-\gamma_k})^M$ is widely accepted. The results of this paper, however, hold not only for this particular choice, but for any efficiency function $f(\cdot)$  that is increasing, S-shaped, approaching unity as $\gamma_k \rightarrow + \infty$, and such that $f(\gamma_k)= o(\gamma_k)$ for vanishing $\gamma_k$.

We consider now several non-cooperative games for utility maximization. First of all, we note that the games reported in \cite{nara2} (optimization with respect to transmit power only), in \cite{meshkati} (optimization with respect to transmit power and linear receiver choice), and in \cite{buzzijsac} (optimization with respect to transmit power, linear receiver choice and spreading code choice) can be extended to the case that a complex signal, as the one in (\ref{eq:r}), is considered.  We do not give here the full details to avoid duplication of these previous results, but simply mention that it can be shown that for all of these games a unique NE point exists, and that, for the case in which also spreading code optimization is performed, the NE point is Pareto-optimal for $K\leq N$. We prefer to give much more details on the interesting case in which WL filtering is used.

Let us thus assume that the decision rule (\ref{eq:decrule2}) is adopted, and consider the problem of utility maximization with respect to the transmit power and the WL receiver choice, i.e.
\beq
\ds  \max_{p_k, \bd_k} u_k \; , \quad
\forall k=1, \ldots, K \; ,
\label{eq:game}
\eeq
with $p_k \leq P_{k, \max}$ the maximum allowed transmit power for the $k$-th user. We have the following result.

\noindent
 {\bf Proposition 1:}
{\em The non-cooperative game defined in (\ref{eq:game}) admits a unique NE point $(p_k^*, \bd_k^*)$,
for $k=1, \ldots, K$, wherein
\begin{itemize}
\item[-]
$\bd^*_k$ is the unique vector, up to a positive multiplicative constant, resulting from the upper $N$-entries of the $k$-th user widely linear receive filter; and
\item[-]
$p_k^*=\min \{\bar{p}_k, P_{k, \max} \}$, with $\bar{p}_k$ the $k$-th user's transmit power such that the $k$-th user's maximum SINR $\gamma_k^*$ equals $\bar{\gamma}$, i.e. the unique solution of the equation $f(\gamma)=\gamma f'(\gamma)$, with $f'(\gamma)$ the derivative of $f(\gamma)$. \\
\end{itemize}
}
\noindent
{\bf Proof:} The proof can be given following the same arguments as in \cite{meshkati}. The full details are however
omitted for the sake of brevity.
\hfill \rule{2mm}{2mm}

Consider finally the case in which also spreading code optimization is performed, i.e. we have
\beq
\ds  \max_{p_k, \bd_k, \bs_k} u_k \; , \quad
\forall k=1, \ldots, K \; .
\label{eq:game2}
\eeq
We now have the following result.

\noindent
 {\bf Proposition 2:}
{\em The non-cooperative game defined in (\ref{eq:game2}) admits a unique NE point $(p_k^*, \bd_k^*, \bs_k^*)$,
for $k=1, \ldots, K$, wherein
\begin{itemize}
\item[-]
$\bs^*_k$ and $\bd^*_k$ are the unique (up to a positive scaling factor for the linear receiver; unicity for the spreading codes set is instead meant with respect to their correlation matrix) $k$-th user's
spreading code and receive filter resulting from the upper $N$-entries of the fixed points of iterations
\[
\left\{\begin{array}{llll}
\bd_{k,a}=\sqrt{2p_k}h_k{\bM_a}^{-1}\bs_{k,a}\; \quad & k=1, \ldots, K \\
\bs_{k,a}=\bd_{k,a}/\|\bd_{k,a}\| \; \quad & k=1, \ldots, K
\end{array}
\right.
\]
Denote by $\gamma_k^*$ the corresponding SINR.
\item[-]
$p_k^*=\min \{\bar{p}_k, P_{k, \max} \}$, with $\bar{p}_k$ the $k$-th user's transmit power such that the $k$-th user's maximum SINR $\gamma_k^*$ equals $\bar{\gamma}$, i.e. the unique solution of the equation $f(\gamma)=\gamma f'(\gamma)$, with $f'(\gamma)$ the derivative of $f(\gamma)$. \\
\end{itemize}
Moreover, for $K \leq 2N$, the NE point is Pareto-optimal.}

\noindent
{\bf Proof:}
Let $\bD=\left[\bd_{1,a}, \ldots ,\bd_{K,a}\right]$, and
$\bX = (\bS, \bD)$, and denote by $\bp_{-k}=[p_1, \ldots, p_{k-1}, p_{k+1}, \ldots, p_K]^T$ the $(K-1)$-dimensional vector containing the transmit powers of all the users except the $k$-th one.
Upon defining
\beq
\ds I_k (\bX, \bp_{-k})= \frac{1}{2h_k^2(\bd_{k,a}^H \bs_{k,a})^2}\left[2\N \|\bd_{k,a}\|^2 + \ds \sum_{i \neq k}
2p_i h_i^2 (\bd_{k,a}^T \bs_{i,a})^2\right] \; ,
\label{eq:Ik_nogamma}
\eeq
the utility function for the considered game can be written in the form
\beq
u_k(p_k, I_k(\bX, \bp_{-k})) \; .
\label{eq:prop_1}
\eeq
Since, for fixed power $p_k$, the utility function is a decreasing function of $I_k$, the utility function (\ref{eq:utility2}) is said to be \emph{separable} in the two parameters, $\bp$ and $\bX$ \cite{sung}, and the corresponding game is a \emph{separable game}\footnote{Otherwise stated, in a separable game the utility function can be written as in (\ref{eq:prop_1}), and, for any fixed $p_k$, the utility function is a decreasing function of $I_k$.}. Let us denote by $\G_X(\bp)$ and  by $\G_p(\bX)$ the subgame arising from utility maximization with respect to the spreading code and uplink linear receiver optimization for a fixed transmit power configuration $\bp$, and the subgame arising from utility maximization with respect to
transmit power for fixed spreading codes and linear receivers, respectively. Based on the results of the previous section it can be shown that the subgame $\G_X(\bp^*)$ admits a  NE point $\bX^* = (\bS^*, \bD^*)$, arising from the  fixed point of the iterations  (\ref{eq:iterazionizz}). Similarly, for any user $k$, it can be shown that the transmit power $p_k^*$, as defined in the text of Proposition 2, is an NE point for the subgame $\G_p(\bS^*, \bD^*)$ \cite{nara2}.
Given any interference $I_k(\bX, \bp_{-k})$, the power $p_k^* = \min (\bar{\gamma} I_k(\bX, \bp_{-k}), P_{k, \max})$ maximizes the utility of the user $k$ and it is a continuous function of the interference $I_k$. Moreover it is easily seen that denoting by $\bX^*(\bp)$ the NE of the subgame $\G_{X}(\bp)$,
 the interference function $I_k(\bX^*(\bp),\bp_{-k})$ in (\ref{eq:Ik_nogamma}) is continuous in $\bp$, for $k = 1, \ldots, K$. According thus to Theorem 2 in \cite{sung} the existence of an NE for the game (\ref{eq:game2}) is guaranteed.
Finally, as regards the fact that the NE is Pareto-optimal for $K\leq 2N$, this immediately descends from the result, shown in the previous section, that for $K\leq 2N$ iterations (\ref{eq:iterazionizz}) converge to a set of orthonormal augmented signatures, thus implying that the users at the NE enjoy separate channels, with no multiuser interference.
\hfill \rule{2mm}{2mm}

\subsection{Network performance prediction through LSA}\label{LSAsec}
Following the approach of \cite{buzzijsac},
we now show how LSA arguments can be used to predict the network performance. In particular, we focus here on predicting
the utility profile of the active users in a large CDMA system (i.e. a CDMA system with $N$ and $K\rightarrow\infty$ but $K/N=\alpha$) employing WL filtering at the receiver, and with no spreading code optimization.
LSA tools were first developed in \cite{Tse-Hanly}, where the following heuristic relation for the SINR achieved by the generic $k$-th user with linear MMSE detection and random unit-norm spreading codes is derived for real channels
\begin{equation}\label{eqSINR_r}
\gamma_k \approx \frac{p_k h_k^2}{N_0/2+\frac{1}{N}\sum_{j\neq k}\frac{p_k h_k^2 p_j h_j^2}{p_k h_k^2+p_j h_j^2\gamma_k}} \; .
\end{equation}
Generalizing (\ref{eqSINR_r}) to the case of a complex channel and WL filtering at the receiver, the SINR $\gamma_k$ achieved by user $k$ can be shown (the proof is omitted for the sake of brevity) to satisfy the equation
\begin{equation}\label{eqSINR}
\gamma_k \approx \frac{2p_k h_k^2}{2N_0+\frac{1}{2N}\sum_{j\neq k}\frac{4p_k h_k^2 p_j h_j^2}{2p_k h_k^2+2p_j h_j^2\gamma_k}} \; .
\end{equation}
The above equation can be used to come up with a non-iterative power control equation for the maximization of the energy-efficiency. Indeed, since all users are to achieve the same target SINR $\bar{\gamma}$, it is reasonable to assume that they must be received with the same power, i.e.
\begin{equation}\label{Pr}
2h_1^2p_1=2h_2^2p_2=\ldots=2h_K^2p_K=P_R \; .
\end{equation}
Substituting (\ref{Pr}) into (\ref{eqSINR}) we obtain
\begin{equation}
P_R=\frac{2N_0\bar{\gamma}}{1-\frac{\bar{\gamma}\alpha}{2(1+\bar{\gamma})}} \; .
\end{equation}
Therefore we can devise the following power control algorithm
\begin{equation}\label{It_Tse_Hanly}
p_k=\min\left(\frac{1}{2h_k^2}\frac{2N_0\bar{\gamma}}{1-\frac{\bar{\gamma}\alpha}{2(1+\bar{\gamma})}},p_{\rm max}\right) \qquad\forall k=1,\ldots,K \, ,
\end{equation}
with $p_{\rm max}$ the maximum allowed transmit power, assumed to be the same for all the users.
Of course, once the transmit powers are known, the actual achieved SINRs can be computed using (\ref{eqSINR}), and the achieved utilities are given by (\ref{eq:utility2}).
Equation (\ref{It_Tse_Hanly}) is the generalization (to the case of WL filtering) of a power control algorithm reported in \cite{Tse-Hanly}. However, it assumes that all users are received with the same power, which is a realistic assumption only if no user ends up transmitting at maximum power. In fact, a user transmitting at maximum power will not achieve the target SINR and therefore (\ref{Pr}) is no longer true. Consequently, as our numerical results will show, (\ref{It_Tse_Hanly}) exhibits a poor prediction accuracy when many users are transmitting at their maximum power. To circumvent this problem, note that in order to carefully predict the users' utility profile, the number of users transmitting at maximum power needs to be estimated.
Denoting by $N_m $ such a number, an estimate of $N_m$ can be obtained according to the following formula:
\begin{equation}
\widehat{N_m}=\sum_{i=1}^K u\left(\frac{1}{2h_k^2}\frac{2N_0\bar{\gamma}}{1-\frac{\bar{\gamma}\alpha}{2(1+\bar{\gamma})}}-p_{\rm max}\right) \;,
\end{equation}
with $u(\cdot)$ denoting the unit-step function.
It is also fairly reasonable to assume that the users transmitting at maximum power will be the $N_m$ users with the smallest modula of the channel coefficients. Therefore, ordering the modula of the channel gains in descending order we can write now the SINR for the generic active user as follows:
\begin{equation}\label{Transmit_Power}
\frac{2NP_k}{4NN_0+\frac{u_1P_k}{1+\bar{\gamma}}+\sum_{i=K-u2+1}^K\frac{2P_kp_{\rm max}h_i^2}{P_k+2p_{\rm max}h_i^2\bar{\gamma}}}=\bar{\gamma} \; ,
\end{equation}
with $P_k=2h_k^2 p_k$.
Solving this equation with respect to $P_k$ gives the desired receive power for user $k$. Actually, since equation (\ref{Transmit_Power}) can be written for any $k=1,\ldots,K$, the solution $P_k$ is the desired receive power for all users, that is $P_R$. Once $P_R$ is known, the powers for all users are easily found as follows:
\begin{equation}\label{ImprovedLSA}
p_k= \min (P_k/2h_k^2,p_{\max}) \; , \, \quad k=1,\ldots,K \; .
\end{equation}
Once the powers are known, the actual achieved SINRs, can be easily computed for all users. Then, the users' utilities are given by
\begin{equation}
v_k=\frac{RL}{M} \frac{f(\gamma_k)}{p_k} \; , \, \quad k=1,\ldots,K \; .
\end{equation}

\section{Numerical results}
In this section we present some simulation results that give insight into the performance of the proposed transceiver optimization algorithm and, also, corroborate, the validity of the theoretical findings.
We consider a DS/CDMA system with BPSK modulation (so that the received signal is improper), and with a randomly generated starting signature set.

In Figs. 1 and 2 we compare the transceiver optimization iterations (\ref{eq:iterazionizz}) with the classical MMSE signature update iteration of \cite{ulukusyates}. The processing gain has been set to $N=15$, and the users' received powers have been randomly generated.
Denote by $\bS_a$, and $\bS$ the sets of augmented and non-augmented signatures, respectively.
Fig. 1 shows the minimum and maximum eigenvalue of the matrices $\bS_a^H\bS_a$, and $\bS^H\bS$ versus the iteration index for two different number of users, namely $K=10$ and $K=20$. As expected, for $K=10$, the maximum and minimum eigenvalues converge to 1 in both cases, implying that both $\bS^H\bS$, and  $\bS_a^H\bS_a$ converge to the identity matrix $\bI_K$, whereas for $K=20$ only $\bS_a^H\bS_a$ converges to $\bI_K$. This can be explained noting that since the matrix $\bS^H\bS$ has rank $N$, it cannot converge to $\bI_K$ when $K>N$, whereas the matrix $\bS_a^H\bS_a$ has rank $2N$, which makes the convergence to $\bI_K$ possible also for $N < K \leq 2N$.
Fig. 2 shows the WL-TWSC and the TWSC of the non-augmented set $\bS$ versus the iteration index for $K=10$ and $K=20$.
Both the WL-TWSC and the TWSC can be lower bounded as in (\ref{Lower}). Given the randomly generated received powers the lower bounds have been found to be
\begin{equation}\label{tab:1}
\begin{tabular}{lc}
\toprule
$\rm{tr}(\bA^2)=5.36$ & for $K=10$\\
$\rm{tr}(\bA^2)=12.08$ & for $K=20$\\
\bottomrule
\end{tabular}
\end{equation}
As expected, the lower bound is achieved by both WL-TWSC and TWSC when $K=10$, but only by WL-TWSC when $K=20$.
These two figures give evidence of the fact that WL filtering coupled with spreading code optimization can double the number of users able to enter the network without suffering any multiuser interference, with respect to previous transceiver optimization algorithms.

Fig. 3 addresses the performance of iterations (\ref{eq:iterazionizz}) when $K>2N$. Here, the processing gain has been set to $N=5$, the number of users to $K=12$, and the users' received powers are
\begin{equation}\label{tab:2}
P_1=11.51 \; , \quad
P_2=7.94 \; , \quad \mbox{and} \quad
P_k=1\; \;    {\rm for} \; \, k=3, \ldots, K \; .
\end{equation}
It is easy to see that the two users with the largest powers are oversized \cite{optimal}. In this scenario, the optimum signature set dedicates each oversized user a signature orthogonal to all other users, while the other users are given generalized Welch-bound signatures. Given the powers in (\ref{tab:2}), the eigenvalues of $\bS_a\bA\bS_a^H$ associated to the optimum set of signatures can be shown to be
\begin{equation}\label{tab:3}
\lambda_1=P_1\; , \quad
\lambda_2=P_2\; , \quad \mbox{and} \quad
\lambda_k=\frac{\sum_{i=3}^K P_i}{2N-2}=1.25\; \;  {\rm for}  \; \, k=3, \ldots, K \; .
\end{equation}
Fig. 3 shows the eigenvalues of $\bS_a\bA\bS_a^H$ versus the iteration index. As expected, the eigenvalues converge to the optimum values, implying that $\bS_a$ converges to the optimum set.

Consider now the system performance of the maximum energy-efficiency non-cooperative game at the NE.
Assume a processing gain $N=11$, and a packet length $M=120$;
for this value of $M$ the equation $f(\gamma)=\gamma f'(\gamma)$ can be shown to admit the solution $\bar{\gamma}=6.689 = 8.25$dB. A single-cell system is considered, wherein users may have random positions with a distance from the AP ranging from 10m to 500m. The channel coefficient $h_ke^{j \phi_k}$ for the generic $k$-th user is assumed to be a zero-mean complex Gaussian random variate with variance equal to $d_k^{-3/2}$, with $d_k$ being the distance of user $k$ from the access point (AP). We take the ambient noise level to be $\N=2.5 \cdot 10^{-10}$W/Hz, while the maximum allowed power $P_{k,\max}$ is $0$dBW. We present the results of averaging over $10^5$ independent realizations for the users locations, fading channel coefficients and starting set of spreading codes. More precisely, for each iteration we randomly generate an $N \times K$-dimensional spreading code matrix with entries in the set $\left\{-1/\sqrt{N}, 1/\sqrt{N}\right\}$; this matrix is then used as the starting point for the games that include spreading code optimization, and as the spreading code matrix for the games that do not perform spreading code optimization. Note that even though the starting spreading codes are real, at the fixed point of iterations (\ref{eq:iterazionizz}) they are complex vectors.

Figs.  4 - 7 show the achieved average utility (measured in bits/Joule), the average user transmit power, the average achieved SINR, and the average fraction of users transmitting at maximum power at the receiver output versus the number of users, for the game in \cite{meshkati}, the game in \cite{nara2}, the game in \cite{buzzijsac}  and for the newly proposed non-cooperative games with WL processing at the receiver. Inspecting the curves,
it is seen that the approach based on WL filtering largely outperforms the games of
\cite{meshkati,nara2,buzzijsac}.
In particular, it is seen that for $K\leq 2N$ a very substantial performance gain can be obtained by resorting to spreading code optimization and WL processing; indeed, when $K\leq 2N$, the use of spreading code optimization coupled with WL filtering permits granting to the active users separate (i.e., orthogonal) channels, so that the multiaccess channel reduces to a superposition of $K$ separate single-user
additive white Gaussian noise (AWGN) channels. Evidence of this is given by the horizontal lines representing the system performance of spreading code optimization plus WL filtering in the range $1<K \leq 2N$.
In Fig. 7 we show the fraction of users transmitting at the maximum power: this is actually the fraction of users not being able at the NE to achieve the target SINR. As expected, the smaller fraction corresponds to the games based on the WL filtering, and also in this case we have an horizontal line representing the performance for $K\leq 2N$.

Figs. 8 compares the performance of the LSA prediction algorithms (\ref{It_Tse_Hanly}) and (\ref{ImprovedLSA}).
The actual users' utility profiles obtained through power control and both WL and linear receiver optimization have been contrasted to the users' utility profiles predicted by (\ref{It_Tse_Hanly}) and (\ref{ImprovedLSA}). It is seen that both with WL and linear filtering at the receiver, the LSA-based algorithm is capable to predict the actual users' utility profile with a much higher accuracy than the algorithm (\ref{It_Tse_Hanly}). In particular, note how the prediction accuracy of (\ref{It_Tse_Hanly}) gets worse as the number of users increases, while the accuracy of the newly proposed algorithm is almost insensitive to the number of users. This can be explained noticing that a larger fraction of users ends up transmitting at maximum power as $K$ increases, and recalling from section \ref{LSAsec} that (\ref{It_Tse_Hanly}) does not take this circumstance into account.

\section{Conclusions}
This paper has considered the issue of non-cooperative transceiver optimization in wireless data networks with WL filtering. Unlike previous work in this area, a baseband complex representation of the data has been considered, and  emphasis has been given to the case in which the received signal is improper, showing that the system performance can be significantly improved by resorting to WL filtering structures. We have solved the problem of non-cooperative spreading code design and WL receiver choice for SINR maximization and/or MSE minimization, and have provided an update algorithm (i.e., iterations (\ref{eq:iterazionizz})) that has one stable fixed point coincident with the global optimum point. Relevant properties of the fixed point of Eq. (\ref{eq:iterazionizz}) have been shown, discussed, and confirmed by numerical simulations. In particular, spreading code optimization coupled with WL filtering has been shown to allow complete suppression of  multiuser interference even in overloaded networks, up to a number of users that is twice the processing gain. It has been also shown that WL filtering brings a substantial improvement to the energy-efficiency of multiuser wireless data networks, and the corresponding system performance at the NE has been analyzed, also through LSA arguments.

\begin{figure}
\centering
\includegraphics[width=12cm]{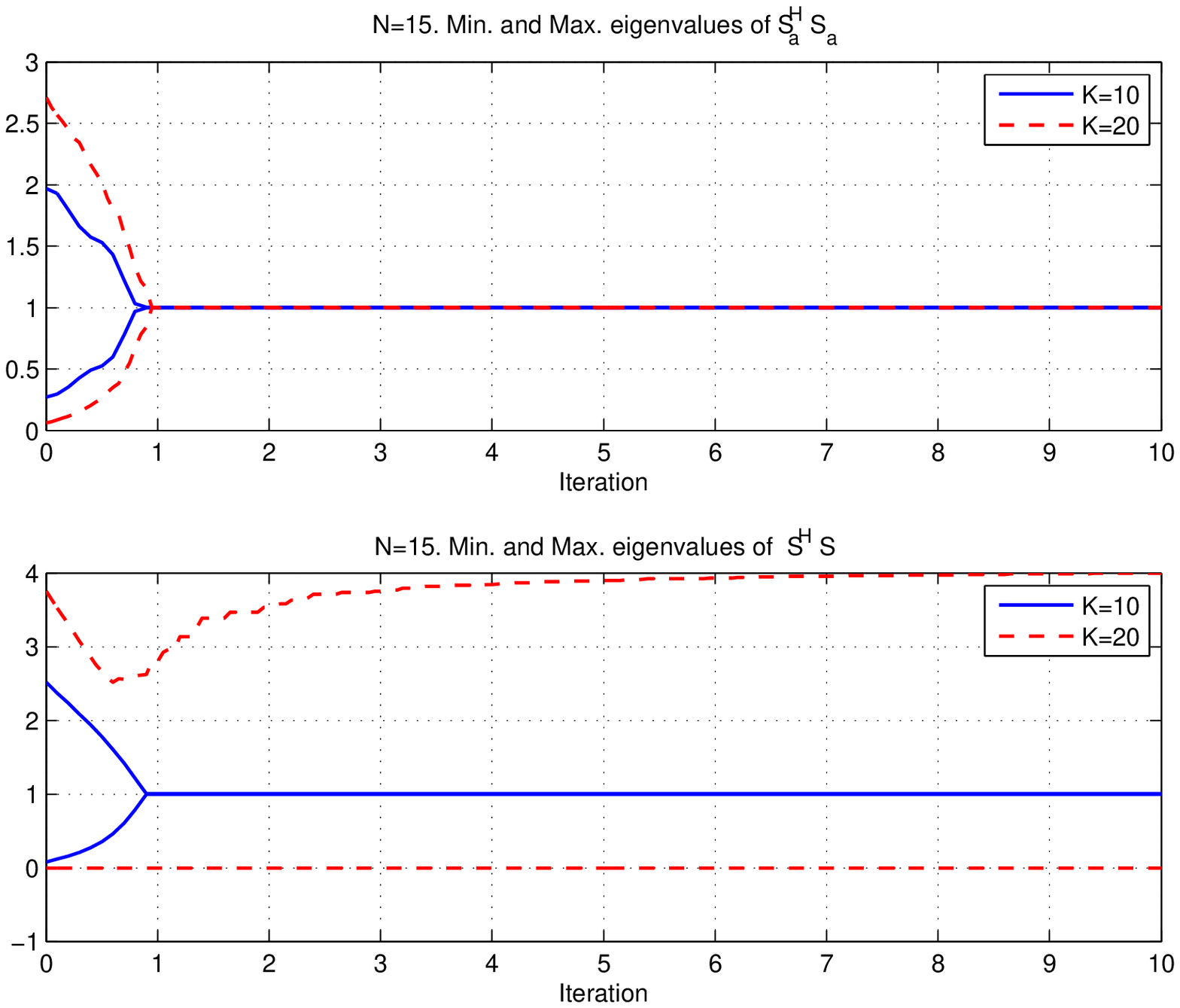}
\caption{Maximum and minimum eigenvalues of $\bS_a^H\bS_a$ and $\bS^H\bS$ versus the iteration index for $K=10,20$. The system processing gain is $N=15$.} \label{fig:1}
\end{figure}

\begin{figure}
\centering
\includegraphics[width=12cm]{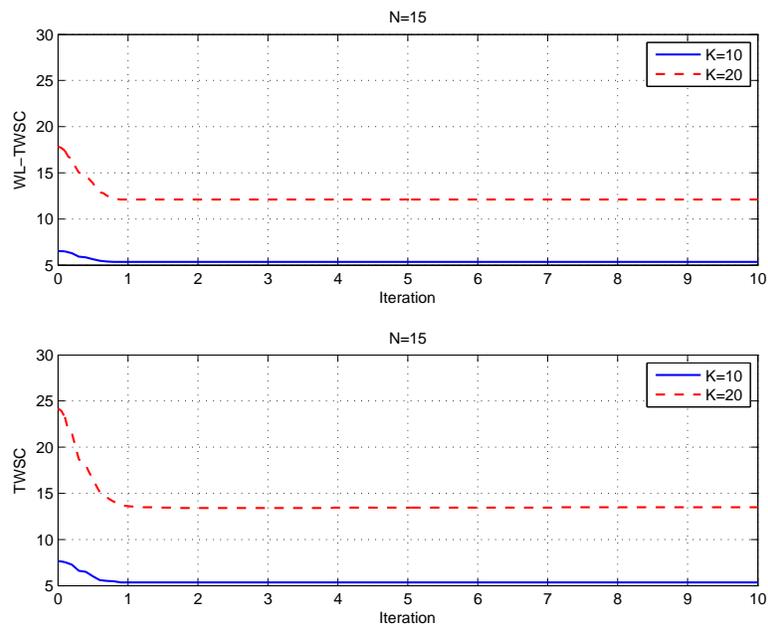}
\caption{WL-TWSC and TWSC versus the iteration index for $K=10,20$. The system processing gain is $N=15$.} \label{fig:2}
\end{figure}

\begin{figure}
\centering
\includegraphics[width=12cm]{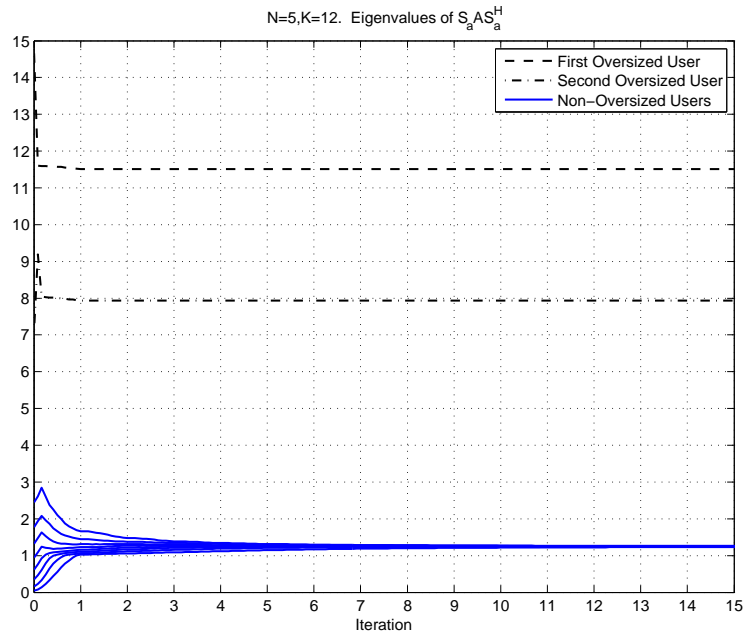}
\caption{Eigenvalues of $\bS_a\bA\bS_a^H$ versus the iteration index for $K=12$. The system processing gain is $N=5$.} \label{fig:3}
\end{figure}

\begin{figure}
\centering
\includegraphics[width=12cm]{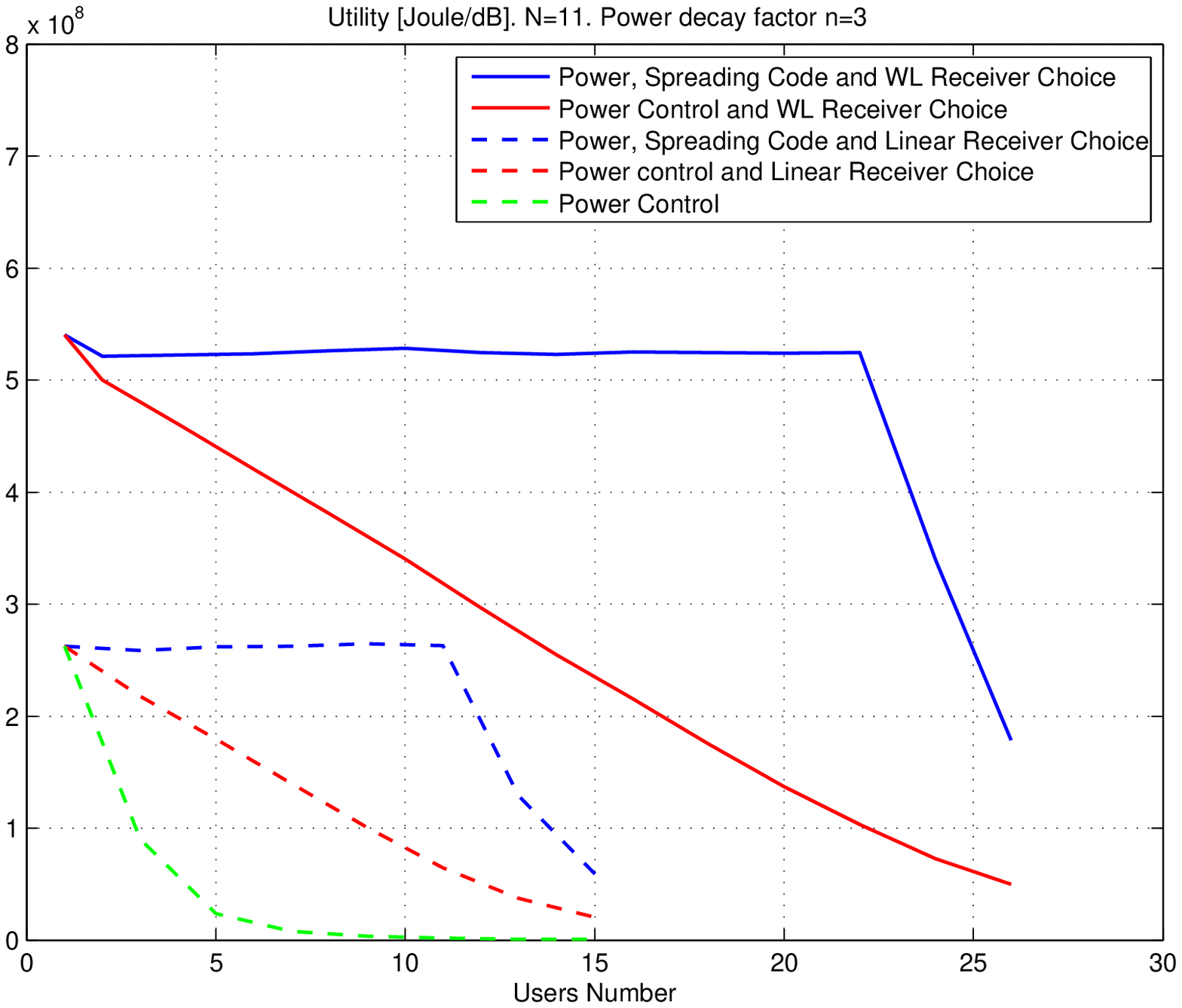}
\caption{Achieved average utility versus number of active users for the maximum energy-efficiency non-cooperative games. The system processing gain is $N=11$.} \label{fig:4}
\end{figure}

\begin{figure}
\centering
\includegraphics[width=12cm]{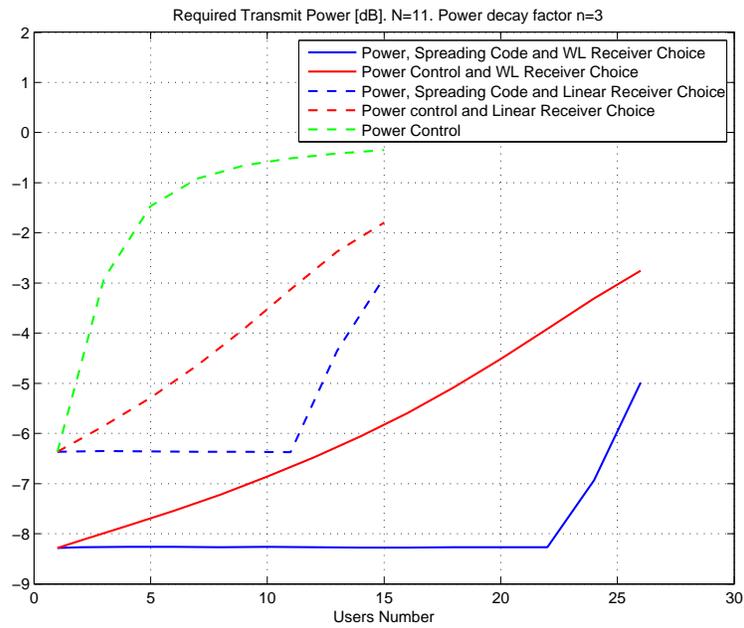}
\caption{Average transmit power versus number of active users for the maximum energy-efficiency non-cooperative games. The system processing gain is $N=11$.} \label{fig:5}
\end{figure}

\begin{figure}
\centering
\includegraphics[width=12cm]{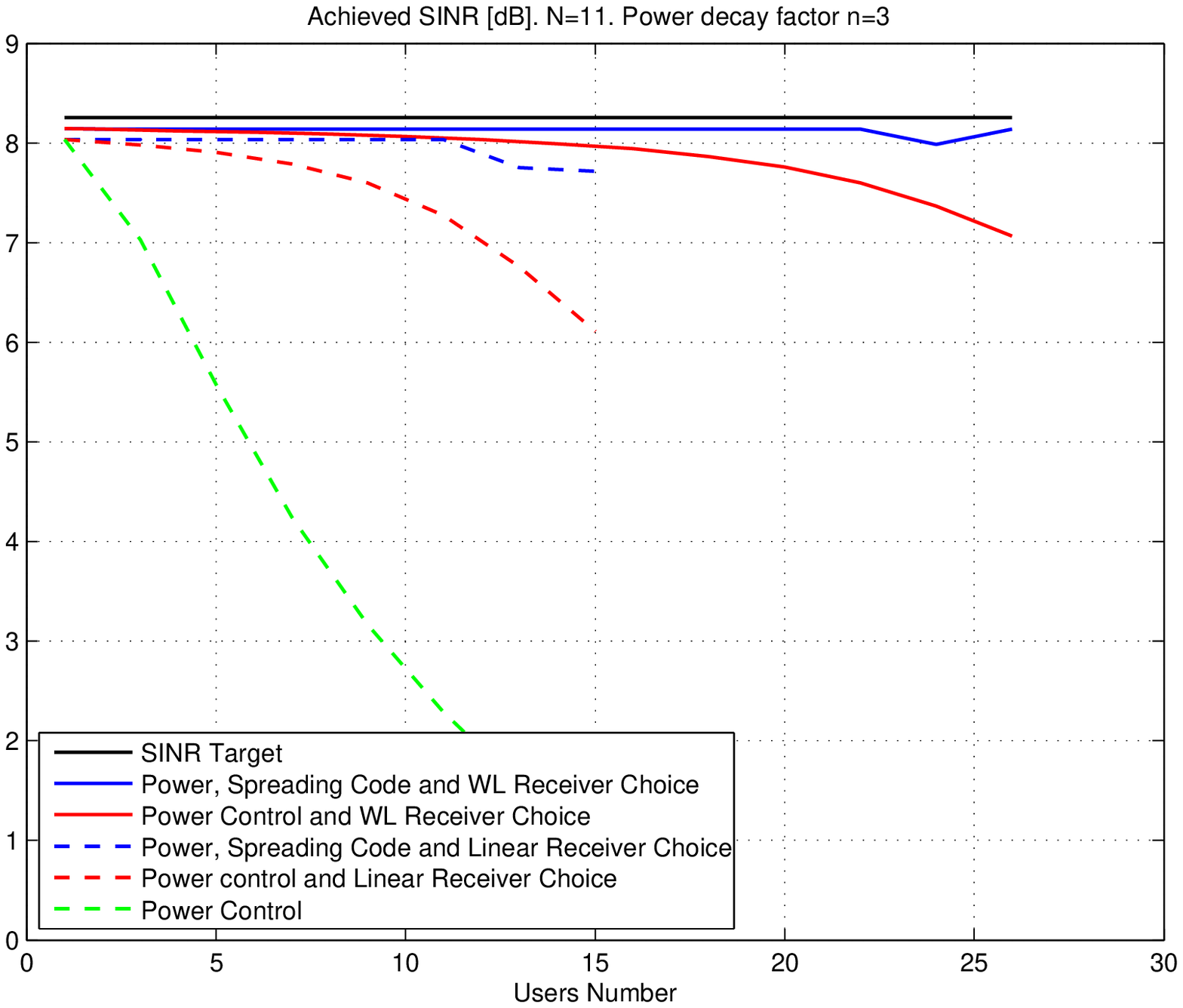}
\caption{Achieved average output SINR versus number of active users for the maximum energy-efficiency non-cooperative games. The system processing gain is $N=11$.} \label{fig:6}
\end{figure}

\begin{figure}
\centering
\includegraphics[width=12cm]{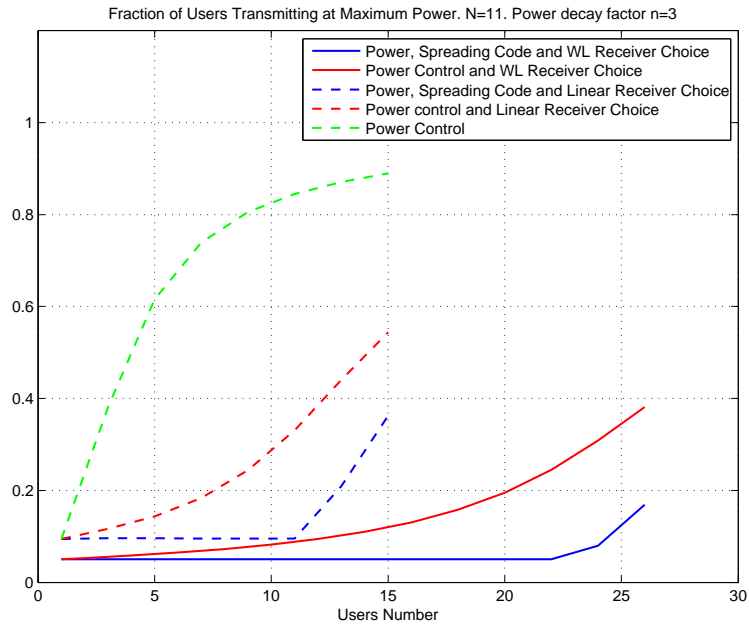}
\caption{Average fraction of users transmitting at their maximum allowed power versus number of active users for the considered non-cooperative games. The system processing gain is $N=11$.} \label{fig:7}
\end{figure}

\begin{figure}
\centering
\includegraphics[width=12cm]{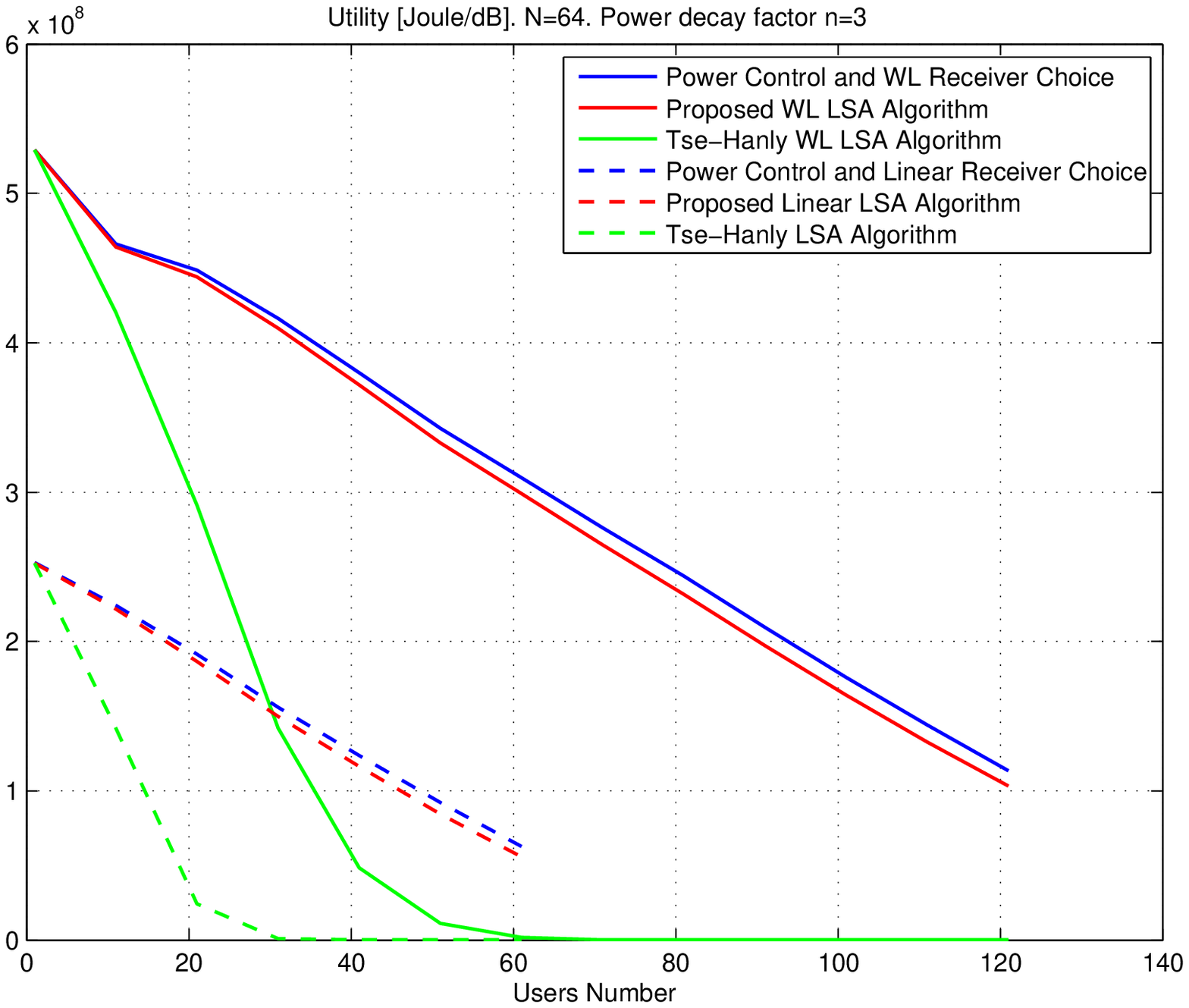}
\caption{LSA-based users' utility profile. The system processing gain is $N=64$. The power decay factor is n=3} \label{fig:8}
\end{figure}


\begin{thebibliography}{99}



\bibitem{rupf}
M. Rupf and J. L. Massey,
``Optimum sequence multisets for synchronous code-division multiple-access channels,''
{\em IEEE Trans. Inf. Theory,} Vol. 40, pp. 1261 - 1266, Jul. 1994.

\bibitem{optimal}
P. Viswanath and V. Anantharam, ``Optimal sequences and sum capacity of synchronous CDMA systems,''
{\em IEEE Trans. Inf. Theory}, Vol. 45, pp. 1984 - 1991, Sept. 1999.

\bibitem{optimal0}
P. Viswanath, V. Anantharam, and D. Tse, ``Optimal sequences, power control and capacity of spread-spectrum
systems with multiuser receivers,''
{\em IEEE Trans. Inf. Theory}, Vol. 45, pp. 1968 - 1983, Sept. 1999.



\bibitem{ulukusyates}
S. Ulukus and R. Yates,
``Iterative construction of optimum signature sequence sets in synchronous CDMA systems,''
{\em IEEE Trans. Inf. Theory}, Vol. 47, pp. 1989 - 1998, July 2001.


\bibitem{ulukusyener}
S. Ulukus and A. Yener,
``Iterative transmitter and receiver optimization for CDMA networks,''
{\em IEEE Trans. Wireless Commun.}, Vol. 3, pp. 1879-1884, Nov. 2004.

\bibitem{ensuring}
P. Anigstein and V. Anantharam, ``Ensuring convergence of the MMSE iteration for interference avoidance to the global optimum,'' {\em IEEE Trans. Inf. Theory}, Vol. 46, pp. 873-885, Sept. 2000.


\bibitem{rose}
C. Rose,
``CDMA codeword optimization: Interference avoidance and convergence via class warfare,'' {\em IEEE Trans. Inf. Theory}, vol. 47, pp. 2368-2382, Sept. 2001.



\bibitem{rose2}
C. Rose, S. Ulukus and R. Yates,
``Wireless systems and interference avoidance,''
{\em IEEE Trans. Wireless Commun.}, Vol. 1, pp. 415 - 428, July 2002.

\bibitem{concha}
J. I. Concha and S. Ulukus,
``Optimization of CDMA signature sequences in multipath channels,''
{\em Proc. of the 53rd IEEE Vehic. Technology Conference},
Rhodes, Greece, May 2001.


\bibitem{popescurose}
D. C. Popescu and C. Rose,
``Codeword optimization for uplink CDMA dispersive channels,''
{\em IEEE Trans. Wireless Commun.}, Vol. 4, pp. 1563 - 1574, July 2005.

\bibitem{honig}
G. S. Rajappan and M. L. Honig, ``Signature sequence adaptation for DS/CDMA with multipath,''
{\em IEEE J. Selected Areas Commun.}, Vol. 20, pp. 384-395, Feb. 2002.




\bibitem{gtbook}
D. Fudenberg and J. Tirole, {\em Game Theory}, Cambridge, MA: MIT Press, 1991.




\bibitem{gt}
A. B. MacKenzie and S. B. Wicker, ``Game theory in communications: Motivation, explanations, and applications to power control,''
{\em Proc. IEEE Global Telecommun. Conference}, San Antonio, TX, 2001.

\bibitem{rodriguez}
V. Rodriguez,
``An analytical foundation for resource management in wireless communication,''
{\em Proc. IEEE Global Telecommun. Conference}, San Francisco, CA, Dec. 2003.

\bibitem{nara2}
C. U. Saraydar, N. B. Mandayam and D. J. Goodman, ``Efficient power control via pricing in wireless data networks,''
{\em IEEE Trans. Commun.}, vol. 50, pp. 291-303, Feb. 2002.

\bibitem{meshkati}
F. Meshkati, H. V. Poor, S. C. Schwartz and N. B. Mandayam,
``An energy-efficient approach to power control and receiver design in wireless data networks,''
{\em IEEE Trans. Commun.}, Vol. 53, pp. 1885-1894, Nov. 2005.



\bibitem{buzzijsac}
S. Buzzi and H.V. Poor,
``Joint receiver and transmitter optimization for energy-efficient CDMA communications,''
{\em IEEE J. Sel. Areas Commun.}, Vol. 26, pp. 459 - 472, Apr. 2008.


\bibitem{bacci}
G. Bacci, M. Luise, H. V. Poor and A. Tulino,
``Energy efficient power control in impulse radio UWB networks,''
{\em IEEE J.  Selected Topics  Sig. Process.}, Vol. 1, pp. 508-520, Oct. 2007.

\bibitem{eusipco2008}
S. Buzzi, H. V. Poor and D. Saturnino,
``Energy-efficient resource allocation in multiuser MIMO systems: A game-theoretic framework,''
{\em Proc. of the 16th European Signal Processing Conference}, Lausanne, Switzerland, August 2008.



\bibitem{buzzieurasip2009}
S. Buzzi, H. V. Poor and D. Saturnino, ``Adaptive cross-layer distributed energy-efficient
resource allocation algorithms for wireless data networks,''
{\em EURASIP Journal on Advances in Sig. Proc., Special Issue on Cross-Layer Design for the Physical, MAC,
and Link Layer in Wireless Systems}, 2009.


\bibitem{Buzzi}
S. Buzzi, M. Lops and A. M. Tulino, ``A generalized
minimum-mean-output-energy strategy for CDMA systems with improper
MAI,'' {\em IEEE Trans. Inform. Theory}, Vol. 48, pp. 761-767,
March 2002.

\bibitem{BuzziLops}
S. Buzzi and M. Lops, ``Performance analysis for the improved
linear multiuser detectors in BPSK-modulated  DS/CDMA systems,''
{\em IEEE Trans. Commun.}, Vol. 51, pp. 37-42, January 2003.

\bibitem{Yoon}
Y. Yoon and H. Leib, ``Maximizing SNR in improper complex noise
and applications to CDMA,'' {\em IEEE Commun. Letters}, Vol. 1,
pp. 5-8, January 1997.

\bibitem{lampe1}
 H. Gerstacker, R. Schober and A. Lampe, ``Receivers with widely-linear processing for frequency-selective channels,'' {\em IEEE
Trans. Commun.}, Vol. 51, pp. 1512-1523, September 2003.

\bibitem{lampe2}
R. Schober, W. H. Gerstacker and L. H.- J. Lampe, ``A blind widely-linear minimum-output-energy algorithm,'' {\em Proc. of the IEEE
Wireless Commun. and Networking Conf. (WCNC 2003)}, New Orleans, LA, Vol. 1, pp.
612-617, March 2003.

\bibitem{Csum}
S. Verd\'u, ``Capacity region of Gaussian CDMA channels: The symbol synchronous case,'' {\em Proc. 24th Annu. Allerton Conf. Communication, Control and Computing}, Monticello, IL, Vol. 1,
pp. 1025-1034, Oct 1986.



\bibitem{sung}
C. W. Sung, K. W. Shum and K. K. Leung,
``Stability of distributed power and signature sequence control for CDMA systems: A game-theoretic framework,"
{\em IEEE Trans. Inf. Theory}, Vol. 52, no. 4, pp. 1775-1780, April 2006.


\bibitem{Tse-Hanly}
D. N. C. Tse, S.V. Hanly ``Linear multiuser receivers: Effective interference, effective bandwidth and user capacity,''
{\em IEEE Trans. Inform. Theory}, vol. 45, pp. 641-657, March. 1999.




\end{thebibliography}
\end{document}